\documentclass[preprint,11pt,authoryear]{elsarticle}

\journal{Journal of Systems and Software}

\usepackage{comment}
\usepackage[most]{tcolorbox} 

\usepackage{amssymb}
\usepackage{amsmath}

\usepackage{multirow}
\usepackage{tcolorbox}
\usepackage{comment}
\usepackage{booktabs}
\usepackage{caption}
\usepackage[margin=1.5cm]{geometry}
\usepackage[hyphens]{url}

\newcommand{\todoc}[2]{{\textcolor{#1}{\textbf{#2}}}}

\newcommand{\todoorange}[1]{\todoc{orange}{\textbf{[[#1]]}}}

\newcommand{\alvine}[1]{\todoorange{Alvine: #1}}

\begin{document}

\begin{frontmatter}



\title{LLMs as Judges: Toward The Automatic Review of GSN-compliant Assurance Cases} 


\fntext[1]{}


\author[inst1]{Gerhard Yu}

\affiliation[inst1]{organization={Lassonde School Of Engineering},
            addressline={York University}, 
            city={Toronto},
            country={Canada}}

\author[inst2]{Mithila Sivakumar}

\author[inst1]{Alvine B. Belle}

\author[inst1]{Soude Ghari}

\author[inst1]{Song Wang}

\author[inst2]{Timothy C. Lethbridge}

\affiliation[inst2]{organization={School of Electrical Engineering and Computer Science},
            addressline={University of Ottawa}, 
            city={Ottawa},
            country={Canada}}

\begin{abstract}
Assurance cases allow verifying the correct implementation of certain non-functional requirements of mission-critical systems, including their safety, security, and reliability. They can be used in the specification of autonomous driving, avionics,  air traffic control, and similar systems. They aim to reduce risks of harm of all kinds including human mortality, environmental damage, and financial loss.
However, assurance cases often tend to be organized as extensive documents spanning hundreds of pages, making their creation, review, and maintenance error-prone, time-consuming, and tedious. Therefore, there is a growing need to leverage (semi-)automated techniques, such as those powered by generative AI and large language models (LLMs), to enhance efficiency, consistency, and accuracy across the entire assurance-case lifecycle. In this paper, we focus on assurance case review, a critical task that ensures the quality of assurance cases and therefore fosters their acceptance by regulatory authorities. We propose a novel approach that leverages the \textit{LLM-as-a-judge} paradigm to automate the review process. Specifically, we propose new predicate-based rules that formalize well-established assurance case review criteria, allowing us to craft LLM prompts tailored to the review task. Our experiments on several state-of-the-art LLMs (GPT-4o, GPT-4.1, DeepSeek-R1, and Gemini 2.0 Flash) show that, while most LLMs yield relatively good review capabilities, DeepSeek-R1 and GPT-4.1 demonstrate superior performance, with DeepSeek-R1 ultimately outperforming GPT-4.1. However, our experimental results also suggest that human reviewers are still needed to refine the reviews LLMs yield.
\end{abstract}




\end{frontmatter}



\section{Introduction}


System assurance is the process of ensuring that a system meets its non-functional requirements (e.g., safety, security, and reliability). For safety-critical systems, this process typically involves the development of assurance cases. An assurance case is a structured, well-reasoned, and auditable collection of arguments, supported by evidence, intended to demonstrate that the system’s non-functional requirements have been correctly and adequately implemented \citep{maksimov2019survey, attwood2015controlled, rinehart2017understanding}. In safety-critical domains, e.g., healthcare, nuclear, and automotive, producers of mission-critical systems use assurance cases to demonstrate compliance with certification standards to regulatory authorities \citep{viger2024ai}. This practice helps prevent system failures and, consequently, catastrophic outcomes such as death, injury, or environmental damage \citep{viger2024ai}. Assurance cases have been employed in safety-critical domains for over fifty years to support certification processes \citep{chowdhury2019criteria}. Their use has also been endorsed by several industry standards, including ISO 26262 \citep{standard201826262}, DO-178C \citep{DO-178c}, and UL 4600 \citep{koopman2023ul}.

A key challenge for creating and maintaining assurance cases is that large numbers of assurance cases tend to be bundled into voluminous documents spanning several hundred pages \citep{maksimov2019survey}. For instance, the safety case for an air traffic control system may exceed 500 pages and reference nearly 400 supporting documents \citep{maksimov2019survey}. The manual creation, refinement, and maintenance of assurance cases are often error-prone, time-consuming, and labor-intensive processes that can take several months, particularly for large, complex, and interconnected systems \citep{graydon2025examining, sivakumar2024prompting, ramakrishna2022automating,  odu2024automatic, nguyen2011experiences}.

The premise for our current research is that relying on (semi-)automated techniques as those enabled by \textbf{Generative AI} through \textbf{LLMs (Large Language Models)} should help tackle these limitations by expediting the execution of system assurance activities. However, existing LLM-based system assurance approaches (e.g., \citep{gohar2024codefeater, shahandashti2024assessing, viger2024ai, sivakumar2024prompting, odu2024automatic, chen2025trusta, murugesan2024automating}) only support a limited set of system assurance activities: assurance case generation, verification, formalization, semantic analysis, assessment, and defeater generation. None of these approaches leverage LLMs to support the automatic \textit{review} of assurance cases. Therefore, the assurance case review mostly remains a manual task that heavily relies on the scrutiny of human reviewers. Yet, review is crucial to guarantee the quality of assurance cases and facilitate their acceptance by regulatory authorities.  Hence, we focus on the review process in this paper.

To tackle this gap and gain a good understanding of the extent to which LLMs can help with the automation of assurance case review, we propose a novel approach that leverages the capabilities of AI models. We created new predicate-based rules that formalize assurance case review. We then used these rules to craft prompts to query the LLM, asking it to review assurance cases automatically. 

We used various state-of-the-art LLMs (GPT-4o, GPT-4.1, DeepSeek-R1, and Gemini 2.0 Flash) to conduct experiments on several assurance cases from various application domains (healthcare, automotive, and computing). To quantitatively assess the LLM-generated reviews, we relied on three metrics that respectively quantify the coherence, usefulness, and informativeness of LLM-generated reviews. To qualitatively assess the LLM-generated reviews, we derived a taxonomy of semantic topics that classifies the review capabilities and issues characterizing the four LLMs. The analysis of our experimental results suggests that LLMs can be effective assistants when reviewing assurance cases, with both DeepSeek-R1 and GPT 4.1 demonstrating superior review capabilities compared to the other LLMs at hand. However, LLMs cannot yet replace human reviewers. This is because, despite being able to detect and describe some quality issues when reviewing assurance cases, LLMs sometimes struggle with some issues, such as hallucination proneness and the tendency to generate reviews that are too generic and therefore not very useful to assurance case developers. It is therefore crucial to keep the human in the loop when completing the LLM-based assurance case review task. 



The contributions of this paper are therefore fourfold:
\begin{itemize} 
    \item  We extract a set of review criteria relevant for the execution of the assurance case review task
     \item  We formalize these review criteria by proposing a set of predicates that allow capturing these review criteria, the associated review issues, and the possible suggestions to resolve these issues.
    \item We used the review criteria formalization to design prompts enabling the LLM-based assurance case review and aligning with the \textit{LLM-as-a-judge} paradigm.
   
    \item We carry out experiments using four state-of-the-art LLMs, and we report the results obtained.

\end{itemize}

The remainder of this paper is organized as follows: Section \ref{sec:background_related-work} describes the background concepts as well as the related work. Section \ref{sec:approach}
describes the LLM-based approach we proposed to automatically support assurance case review. Section \ref{sec:experimental_setup} describes the experimental setups. Section \ref{sec: results} presents the results we obtained from our experiments. Section \ref{sec:threats} describes the threats that could affect our study. We conclude and outline some future work in Section \ref{sec:conclusion}.

\section{Background and Related Work}
\label{sec:background_related-work}

\subsection{Background} 

\subsubsection{What is an assurance case?}
An assurance case is a structured argument, supported by evidence, and intended to demonstrate that a system is acceptable for its intended use with respect to specific non-functional requirements (e.g., safety, security, reliability) \citep{rinehart2017understanding, mansourov2010system}.  Such cases are structured as a hierarchy of claims,
with lower-level claims drawing on concrete evidence, and serving as evidence to justify claims higher in the hierarchy \citep{foster2021integration}. The ultimate goal of an assurance case is to justify its top claim (i.e., root), such that the system supports a critical non-functional requirement \citep{foster2021integration}. There are different types of assurance cases depending on the non-functional requirement they target. For instance, security cases \citep{alexander2011security,mohamad2021security}, safety cases \citep{palin2010assurance} and reliability cases \citep{zhu2018software} are assurance cases focusing on security, safety, and reliability respectively.

Assurance cases support the verification of the correct implementation of system requirements.  They allow engineers to determine which analyses they should perform to verify that correctness and which supporting evidence and rationales they should use to prove these analyses are sufficient to meet the so-called requirements \citep{rao2019comparative}. For a given system, the supporting evidence may include concrete facts such as system specifications, test results, formal reviews,  simulations, as well as resource diagrams \citep{agrawal2019leveraging, mansourov2010system, delavara2023assurance}.
 Assurance cases therefore play a critical role in helping prevent system failure, which could otherwise lead to fatalities, serious injuries, economic losses, or environmental damage \citep{rushby2017assurance}.  Industry standards that promote assurance cases have therefore become more common, 
 including ISO 26262 \citep{standard201826262}, DO-178c \citep{DO-178c}, ISO/PAS 8800 \citep{ISOPAS8800}, and UL4600 \citep{koopman2023ul}. Various regulatory authorities also promote the use of assurance cases to support the certification of systems in compliance with industry standards. These include FDA (Food and Drug Administration) \citep{sujan2021safety, finnegan2014security} and NHTSA (National Highway Traffic Safety Administration).  Such cases are widely applied across multiple domains—including aerospace, railways, automotive, and healthcare—to justify the safe and reliable deployment of mission-critical systems \citep{maksimov2019survey, rushby2017assurance, duan2017reasoning, graydon2017investigation}.

\subsubsection{How to represent assurance cases?} 

Several notations have been developed to represent assurance cases, including textual and graphical formats. Textual notations include traditional prose (i.e. plain prose) and structured prose (also called semi-structured text) that is more formal \citep{holloway2008safety, sivakumar2024prompting}.  
Representing assurance cases in textual formats, especially traditional prose, is challenging due to the use of unclear and poorly structured English, ambiguity, the usage of multiple cross-references that disrupt the flow of the arguments, and the lack of formalism \citep{kelly2004systematic, odu2024automatic}. Graphical notations address these issues.

Several graphical notations allow representing assurance cases. These  notations allow the explicit representation of assurance case elements as well as the relationships between them \citep{wei2019model}. Graphical notations include the Goal Structuring Notation (GSN) \citep{gsnstandard}, Claims-Arguments-Evidence (CAE) \citep{CAE}, and Eliminative Argumentation (EA) \citep{goodenough2015eliminative}. 

To foster standardization and interoperability, the Object Management Group (OMG) proposed the Structured Assurance Case Metamodel (SACM) to support the representation of assurance cases \citep{SACMv2.3, wei2019model}. To model assurance cases, SACM uses three core
concepts \citep{SACMv2.3, mansourov2010system}: 1) \textbf{A claim}: it is expressed as a proposition and presents a statement that is either true or false, and which describes what the assurance case is trying to justify; 2) \textbf{Some evidence}: this is constituted from the system’s available concrete facts and serves as a proof to back the claim; and 3) \textbf{A well-structured argument}: it links the evidence to the claim in a manner that supports the belief that the evidence
supports the claim. 

In this paper, we focus on GSN because it is  the most used graphical notation \citep{shahandashti2024prisma, carlan2016integrated}. An Assurance case represented in GSN is a diagram having a tree-like structure called \textit{goal structure}. GSN diagrams align with the SACM standard. The following core GSN elements allow representing an assurance case in GSN \citep{gsnstandard}: 
\begin{itemize}
    \item \textbf{Goal}: it represents a claim. Usually denoted as \textit{G}, a goal is depicted as a rectangle. 
    \item  \textbf{Strategy}: it represents an argument. It embodies the inference rules that allow
inferring a claim from sub-claims. It is represented as a parallelogram, and denoted as \textit{S}.
    \item \textbf{Solution}: It represents an evidence. Usually denoted as \textit{Sn}, a solution acts as the supporting evidence for a claim. It is represented as a circle.
\end{itemize}

GSN diagrams may also include other GSN elements such as \textit{Context}, \textit{Assumption}, and  \textit{Justification} \citep{gsnstandard}.

The most common relationships between GSN elements are \textit{SupportedBy} and \textit{InContextOf} \citep{gsnstandard}. \textit{SupportedBy} is represented by an arrow with a black head. It represents inferential or evidential relationships between GSN elements \citep{gsnstandard}. \textit{InContextOf} is represented by an arrow with a white head. It represents contextual relationships between GSN elements \citep{gsnstandard}. 

GSN also allows adding decorators to GSN elements. GSN decorators include the \textit{Undeveloped} decorator that allows specifying that a GSN element is not yet fully developed  \citep{gsnstandard}. It is represented by a hollow diamond that is placed at the bottom center side of the GSN element. This decorator is usually applied on a Goal or a  Strategy. 

\begin{figure}
\centering
\includegraphics[width=1.0\linewidth]{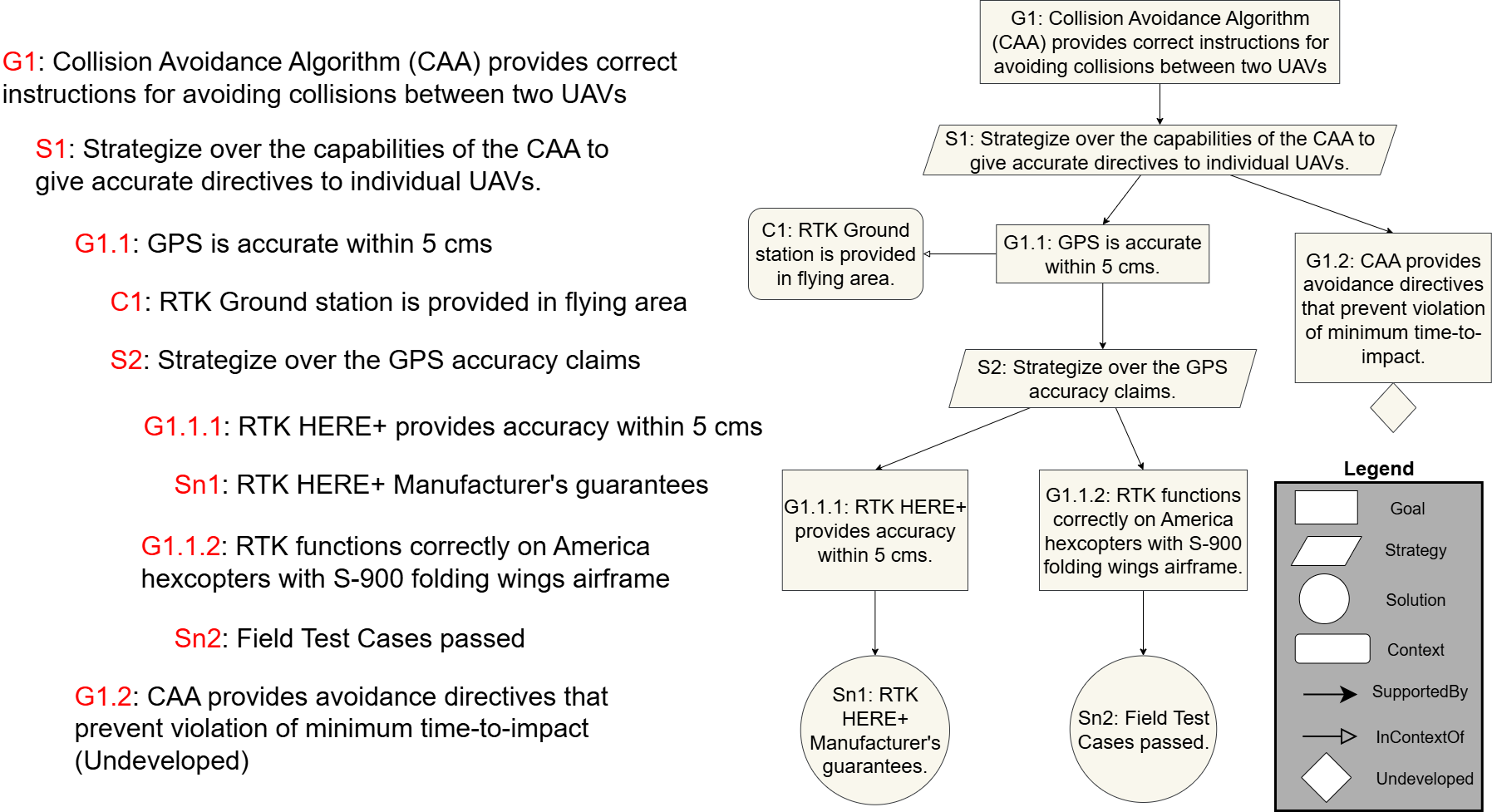}
\caption{On the right,  a partial safety case (GSN diagram) adapted from \cite{vierhauser2019interlocking}; on the left, the equivalent of this safety case in the structured prose 
} 
\label{fig:partial_safety_case}
\end{figure}

Figure \ref{fig:partial_safety_case} depicts a partial assurance case represented using GSN. This figure also shows the equivalent of this assurance case in the structured prose.

\subsubsection{Dialectic Extension of GSN}
\label{subsubsec:dialectic-extension}

The GSN standard \citep{gsnstandard} comprises several extensions that enrich the notations it supports as well as their semantics. These extensions include the argument pattern extension, the dialectic extension, and the confidence argument extension. Also known as the "\textit{investigation of truth}", the dialectic process, is used in assurance cases to strengthen arguments by supporting their constructive criticism and logical dispute. The dialectic extension of GSN allows supporting the dialectic process by relying on a notation that supports various former and new GSN concepts: Goals, Solutions, Challenges, and Defeated. Goals are used to "\textit{challenge a part of an argument}". Solutions are used as a point of reference or evidence to Goals. Challenges are used to indicate that an argument is trying to challenge another different argument. Finally, Defeated  indicates that the relationship of two arguments has been nullified. There are two different kinds of defeaters present in the dialectic extension of the GSN standard: \textbf{Rebuttal} and \textbf{Undercutting} defeaters. A Rebuttal defeater is a supported counter-argument that challenges all or some parts of another argument. An Undercutting defeater refers to additional facts that can be used to challenge parts or all of an argument. 
In this paper, we rely on the dialectic extension of GSN to represent and reason about defeaters. 

\subsubsection{Assurance case review} 
The creation of assurance cases is often a manual process carried out by assurance case developers who have several years of experience in the field. The creation of an assurance case is therefore a subjective task. 
This calls for a rigorous assurance case review process \citep{kelly2004systematic}. The review process helps verify the correct use of syntax rules and establish the traceability among artifacts and the assurance case at hand \citep{yamamoto2016system}. The review process also helps detect potential assurance weakeners (i.e. assurance deficits  and logical fallacies) \citep{shahandashti2024prisma} in arguments early. This informs the improvement of the assurance cases content, yields assurance cases of better quality, and therefore fosters their acceptance by regulatory bodies \citep{kelly2007reviewing}. 

The assurance case review is usually completed before the system is approved for use \citep{kelly2007reviewing}.  \cite{kelly2007reviewing} explains that the assurance case review is usually performed by two parties. One party (i.e. the organization developing the system under scrutiny) has the responsibility to prepare and self-review the assurance case, while the other party (i.e. the certification authority) is responsible for accepting the assurance case. The certification authority considers that an assurance case is acceptable if that assurance case does not comprise major flaws in reasoning or weaknesses in evidence \citep{kelly2007reviewing}.

 Although assurance case review is crucial, there are many issues a reviewer may encounter when manually reviewing an assurance case. Such issues include the need for reviewers to carry out \textit{industrial archaeology} to uncover the assurance case arguments and evidence \citep{gsnstandard, kelly2007reviewing}. Conducting \textit{Industrial archaeology}  is challenging \citep{gsnstandard, kelly2007reviewing}, especially for large assurance cases spanning hundreds of pages. It may drive reviewers to perform superficial rounds of reviews mainly focusing on the presentation of the assurance case, rather than the structure or content of the argument \citep{gsnstandard, kelly2007reviewing}. Once reviewers have uncovered arguments, other issues may arise. These usually relate to the assumptions an assurance case author may sometimes make. More specifically, the author of an assurance case may assume that the reviewers of a given system's assurance case will have as much knowledge and context information about the analyzed system as the developers of the system. However, in practice, most reviewers have a limited knowledge of the system at hand compared to the developers of the systems, and may therefore be confused when reviewing the assurance case \citep{gsnstandard, kelly2007reviewing}. Other issues are inherent to the nature of the review process itself. For instance, the review of assurance cases is usually done manually, which makes it a labor-intensive and very time-consuming task. Furthermore, reviewing an assurance case requires considerable expertise \citep{kelly2007reviewing}.  
 
  
These issues call for the definition of an assurance case review process that follows a clear, systematic, expert-informed, (semi-) automatic, and rigorous methodology.

\subsubsection{What is a Large Language Model (LLM)?} 
LLMs are advanced AI systems with massive parameter sizes,  and trained on large amounts of data coming from diverse sources \citep{chang2024survey, zhao2023survey, hou2024large}. LLMs can emulate human linguistic capabilities and generate plausible texts \citep{hou2024large}. They can therefore summarize text, translate text, and answer questions with human-level language fluency \citep{hou2024large}. 


Prompt engineering supports the optimization of LLM-generated responses through the use of various prompting strategies that guide the LLM with specific instructions on how to perform a given task \citep{mesko2023prompt}. 
 The most common prompting strategies include Zero-Shot prompting, One-Shot prompting, Few-Shot prompting, and Chain-of-Thought (CoT) prompting \citep{choi2025survey, sahoo2024systematic, brown2020language}. Zero-shot prompting is a technique that consists of prompting the LLM to carry out a given task without any explicit training on that task. With that technique, the LLM only relies on the supplied prompt provided during inference
to generate the expected response \citep{hou2024large}. One-shot consists in feeding the LLM at hand with a single example with the aim of improving its learning capabilities \citep{hou2024large}. Few-shot prompting consists of prompting the LLM with a set of examples on how to perform a given task \citep{hou2024large}. The LLM learns from these examples \citep{hou2024large}, which allows improving its reasoning capabilities. CoT is a prompting technique that consists of supplying the LLM with a prompt that breaks down a complex task to perform into simpler ones, which improves the reasoning capabilities of the LLM and allows it to generate well-structured and meaningful responses \citep{wei2022chain, hou2024large}. 

LLMs have become very popular in the Software Engineering field, where they are used to automate various tasks: code review (e.g., \citep{yu2024fine}), software design (e.g., \citep{wan2024software}), unit test generation (e.g., \citep{chen2024chatunitest}), and program repair (e.g., \citep{bouzenia2024repairagent, yin2024thinkrepair}). They have also been used as judges (i.e. raters) to assess the quality of software artifacts \citep{baltes2025evaluation} (e.g., requirements statements \citep{lubos2024leveraging}, acceptance criteria \citep{wang2025multi}). Accordingly, in this paper, we leverage the \textit{LLM-as-a-judge} paradigm \citep{gu2024survey, szymanski2025limitations, shankar2024validates}  and therefore use LLMs as judges to assess the quality of assurance cases during the review process. By merging the scalability of automatic assessment methods with the detailed and context-sensitive reasoning characterizing human expert judgments, this novel paradigm offers a valuable assessment solution to complex and open-ended evaluation problems \citep{gu2024survey}.


\subsection{Related Work}


\subsubsection{Approaches proposed to  review assurance cases} 
\label{subsec:related-work-review-ac}
Several approaches have been proposed to manually review assurance cases (e.g., \citep{kelly2007reviewing, yamamoto2016system, chowdhury2019criteria, rushby2015understanding}. One such approach is the four-stage review process proposed by \cite{kelly2007reviewing}, and currently used by the GSN standard. This approach consists of Argument Comprehension, Well-Formedness (Syntax) Checks, Expressive Sufficiency Checks, and Argument Criticism and Defeat. Another approach is the one by  \cite{yamamoto2016system}. This approach also consists in performing a four-stage review process. The stages comprised in that process are the following: context understanding, problem identification, cause analysis, and revision. \cite{chowdhury2019criteria} have also proposed two criteria to evaluate assurance cases. These criteria respectively assess the structure/notation and the content of the assurance case under review. \cite{rushby2015understanding}  proposed a two-part process that allows reviewing assurance cases using epistemic methods and logic.

A few approaches support the automated review of assurance cases. For instance, \cite{denney2012advocate} proposed an automated review approach supported by AdvoCATE. The latter allows reviewing assurance cases by relying on the safety metrics generated by AdvoCATE. More recently,  \cite{muram2023attest} introduced an automated approach supported by ATTEST. The latter is a Natural Language Processing (NLP) framework that uses a four-stage review process review assurance cases. This process consists of the following stages: text preprocessing, pattern matching, information analysis, and update of assurance cases. 

However, existing approaches proposed to review assurance cases have several limitations. They are usually manual and therefore labor-intensive, time-consuming, and error-prone. The few automatic ones are limited by their supporting tools and their restrictive metamodels \citep{muram2023attest}. Another limitation of some of the existing approaches is their tendency to turn the review process into \textit{industrial archaeology} \citep{kelly2007reviewing, gsnstandard}, which may make them unable to detect potential deficiencies in the assurance case under review. Furthermore, despite the wide adoption of LLMs to support various Software Engineering tasks, none of the existing assurance review approaches has explored the use of LLMs to automate assurance case review. We fill this gap in this paper by exploring the use of LLMs to automate the assurance case review task.

\subsubsection{LLM-based System Assurance Approaches} 
Several approaches (e.g., \citep{fujiwara2025llm, odu2025llm, odu2024automatic, sivakumar2024prompting, sivakumar2024exploring, odu2025smartgsn, viger2024ai, gohar2024codefeater, shahandashti2024assessing, murugesan2024automating,  ghahremani2024avoiding}) rely on LLMs to support system assurance activities. These approaches usually rely on the following prompting strategies: zero-shot, one-shot,
role-based system prompting, and Chain-of-Thought. Most of these approaches focus on two system assurance activities: assurance case creation (e.g., \citep{ sivakumar2024prompting, odu2024automatic, odu2025smartgsn, sivakumar2024exploring, odu2025llm}) and defeater identification (e.g., \cite{viger2024ai, gohar2024codefeater, shahandashti2024assessing}). A few of these approaches (e.g., \citep{chen2025trusta, varadarajan2024enabling, murugesan2024automating, ghahremani2024avoiding}) support some of the following system assurance activities: assurance case verification, formalization, semantic
analysis, and assessment. 

Most of the approaches focusing on assurance case creation rely on GSN to represent assurance cases, and on LLMs from the GPT series to automate that task. For instance, \cite{sivakumar2024exploring, sivakumar2024prompting} relied on GPT-4 -a popular LLM from the GPT series- to automate the generation of safety cases. \cite{odu2024automatic} also relied on two LLMs (i.e. GPT-4 Turbo and GPT-4o) from the GPT series to automatically generate assurance cases from assurance case patterns formalized using predicate-based rules. In subsequent works, Odu et al. automated that approach as an online tool called \textit{SmartGSN} \citep{odu2025smartgsn}, and applied that approach to the automotive sector \citep{odu2025llm}. \cite{fujiwara2025llm}  also proposed an automated method that not only enables the generation of assurance cases using GPT but also allows the generation of mitigations for threats against AI systems.

Most of the approaches (e.g., \citep{viger2024ai, shahandashti2024assessing}) leveraging LLMs to identify defeaters relied on EA to reason about defeaters. Only one approach (i.e. \citep{gohar2024codefeater}) relied on an agnostic notation. Each of these approaches relied on LLMs from the GPT series to automate the defeater identification task. These approaches relied on various metrics to assess their results, including informativeness, usefulness, coherence, and reasonability. \cite{viger2024ai}'s work also proposed a taxonomy of semantic topics resulting from the incremental classification of LLM-generated defeaters.  \cite{shahandashti2024assessing} proposed to rely on a set of predicate-based rules that formalize  EA elements and their connections to prompt GPT-4 Turbo to identify defeaters.

More recently, \cite{graydon2025examining} engaged in a critical review of existing work to determine whether LLMs actually have the capabilities to automatically perform system assurance activities. They concluded that further research still needs to be conducted before a definitive answer is provided. In line with these conclusions, \cite{diemert2025balancing} reviewed existing literature on LLM-based system assurance and identified four common use cases. They also introduced seven novel use cases. Finally, they presented a method for evaluating the risk-benefit tradeoff of a given LLM use case.

However, despite the increasing use of LLMs to support the automation of system assurance activities, none of the existing LLM-based system assurance approaches has explored the use of LLMs to automate the assurance case review task. We address this gap in our paper, and similar to \cite{shahandashti2024assessing} and \cite{odu2024automatic}, we rely on predicate-based rules to support this system assurance task.

\subsubsection{Automatic code review using Large Language Models} 
As \cite{carlan2016integrated} point out, code review is a critical activity in the software system lifecycle. It consists in systematically examining the source code of a system to detect faults overlooked during its development. This allows improving the overall quality of the software under development. There is a rich body of work  (e.g., \citep{yu2024fine, rasheed2024ai, sun2025bitsai, tanleveraging}) that focused on the use of LLMs to automate various Software Engineering tasks such as code review. 
For instance, \cite{yu2024fine} constructed \textit{Carllm} (Comprehensibility of Automated Code Review using Large Language Models), an LLM that is specifically fine-tuned to support automated code reviews. \textit{Carllm} uses CoT prompts to support code review through four main steps that foster the generation of very comprehensive and informative reviews:  issue detection, issue localization, issue explanation, and suggestion of the review fix. These prompts leverage a set of mathematical symbols and notations that allow formalizing code issues. 

In this paper, we adapt \cite{yu2024fine}'s work to propose a novel LLM-based approach that supports the automatic review of assurance cases.

\section{A novel LLM-based Approach to Automatically Review Assurance Cases} \label{sec:approach}

\begin{figure}[h!]
\centering
\includegraphics[width=0.70\linewidth]{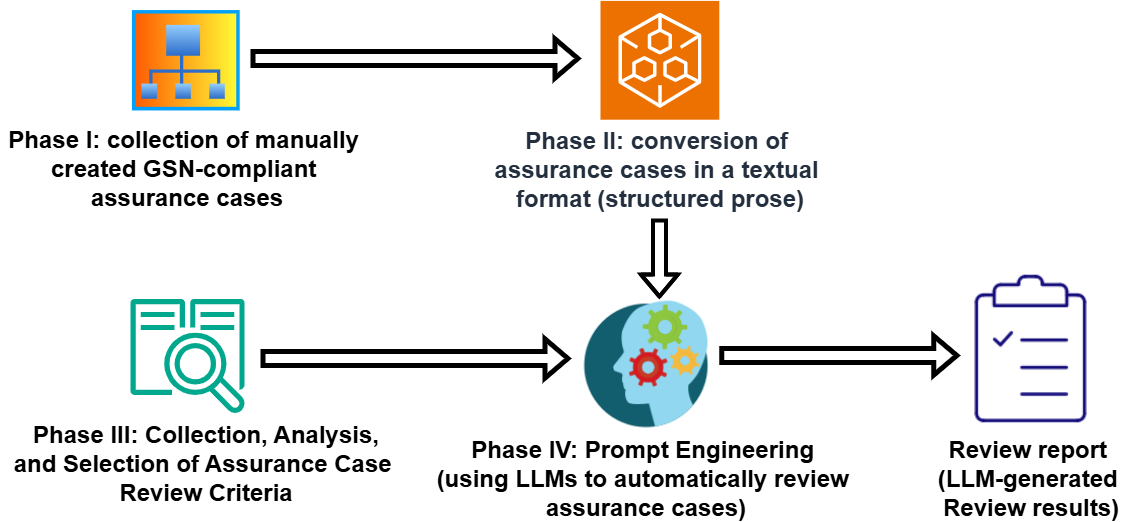}
\caption{High-Level Overview of our LLM-based Assurance Case Review Approach}
\label{fig:approach}
\end{figure}

We propose a novel LLM-based approach to automatically support assurance case review. For this purpose, we develop novel predicate-based rules that formalize assurance case review criteria and issues by tailoring \cite{yu2024fine} et al.'s mathematical symbols and notations to the assurance case review task. Then, we rely on these rules to design CoT prompts tailored to the assurance case review process. These CoT prompts allow us to detect potential issues in assurance cases and suggest solutions to fix them.  Figure \ref{fig:approach} illustrates the various phases of our novel approach. We further describe each of these phases in the remainder of this section.

\subsection{Phase I: Collection of Assurance Cases} 
\label{subsec:data-collection}
In this phase, we followed a systematic approach to select assurance cases, using Google Scholar to identify peer-reviewed papers on system assurance and those offering publicly accessible assurance cases. To foster generalization, we selected a diverse range of assurance cases spanning several application domains. We only retained assurance cases complying with GSN and manually created by assurance case developers.


\subsection{Phase II: conversion of the assurance cases in a textual format (structured prose)} 
In this phase, we converted the assurance cases collected in Phase 1 from GSN into structured prose. This format enables LLMs to interpret and process the safety cases more effectively during the review stage. We manually converted only those assurance cases that did not have a publicly available structured prose version.

\subsection{Phase III: Collection,  Analysis, and Selection of Assurance Case Review Criteria}
Several key studies (e.g., \cite{kelly2007reviewing, yamamoto2016system, rushby2015understanding, muram2023attest}) examine the argument review process (see Section \ref{subsec:related-work-review-ac}). We collected and analyzed these studies, ultimately focusing on the one describing the GSN standard , as it is widely recognized in the system assurance field. The GSN standard (i.e. \citep{gsnstandard}) builds on the work of \cite{kelly2007reviewing} to propose the following four argument review criteria to manually review an argument from an assurance case :

\begin{enumerate}
    \item \textbf{Argument Comprehension}:
    This criterion focuses on identifying the links of each argument in the assurance case; these links focus on finding each claim and its supporting evidence. When using this criterion to review the assurance case, the reviewer must be able to identify key points of an argument, which may include but are not limited to claims, strategies, context of the argument, and evidence of why the argument is valid. The goal is to be able to link each argument with its respective evidence, and if there are unclear links, further annotating the document is necessary to ensure the reviewer understands each connection and how it is supported by their evidence.
    
    \item \textbf{Well-formedness (Syntax)}: This review criteria is concerned with the detection of structural errors in the argument under review. For instance, the presence of \textit{circular arguments} -- arguments in which argument’s premises depend on the argument’s conclusions – are usually not tolerated. This review criteria is also concerned with the identification of claims for which no supporting argument or evidence has been presented, and the identification of pieces of evidence whose role in the argument is unclear. 
    
    \item \textbf{Expressive Sufficiency}:
    The purpose of the expressive sufficiency criterion is to ensure that assurance case arguments are represented in a way that is complete, unambiguous, and understandable to reviewers. This criterion also emphasizes the importance of accompanying documentation to clarify the meaning of each context element thereby preventing misinterpretation and strengthening the overall comprehensibility of the argument. 
    
    \item \textbf{ Argument Criticism and Defeat}:
    This review criterion is concerned with the establishment of arguments overall sufficiency. Hence, it focuses on determining if an argument’s premises, when considered together, are convincing enough to support the argument’s conclusions. The sufficiency of the relationship between an argument’s premises and conclusions can be evaluated by relying on various attributes that are further explained in the GSN standard \cite{gsnstandard}: \textit{coverage}, \textit{dependency}, \textit{definition}, \textit{directness}, \textit{relevance}, and \textit{robustness}. With this review criterion, arguments considered insufficient may be challenged by identifying defeaters that defeat (i.e. undermine) them. We rely on the dialectic extension of GSN to conceptualize these defeaters.

\end{enumerate}

In this paper, we rely on the aforementioned review criteria to review assurance cases. In the GSN standard, these criteria are listed in order of necessity (e.g., we can only check the well-formedness of an argument after checking that its structure is fully 
comprehended) and in an order of increasing difficulty.  This means the latter review criterion requires more significant intellectual effort and domain knowledge than the former does. 

\subsection{Phase IV: Prompt Engineering}
\label{subsec:prompt-engineering}
In this phase, we rely on various LLMs to automatically review manually created assurance cases represented in the structured prose format. For this purpose, we design a set of prompt templates that leverage the formalization of the review criteria identified in Phase III. To instruct LLMs to review assurance cases, we then derive prompts from these prompt templates.

\subsubsection{Description of prompting strategies}
We applied a range of prompt engineering techniques derived from well-established prompting strategies \citep{wei2022chain, kojima2022large} to guide our LLMs. Specifically, we relied on three prompting techniques: 1) zero-shot prompting; 2) zero-shot prompting with CoT; and 3) one-shot prompting with CoT. These variations allowed us to determine how different prompting styles influenced the model’s ability to interpret and reason over the assurance cases.
In the remainder of this paper, we explain the design of the aforementioned prompting techniques.

\subsubsection{Types of LLM Prompts}
In this paper, we rely on three types of prompts: the system prompt, the user prompt, and the assistant prompt.
 As we explained in \citep{odu2024automatic}, we can categorize these three types of prompts into two main groups:
 \begin{itemize}
    
     \item \textbf{The Input passed to the LLM}: This is the LLM prompt. The LLM prompt consists of two parts: the system prompt and the user prompt. The system prompt provides developer-defined instructions that shape how the LLM should behave and what knowledge it should prioritize. The user prompt, in contrast, comes from the end user (e.g., a customer or guest) and represents the direct query to the model. In the sections below, we describe the system and user prompts we designed to query the LLMs for generating assurance case reviews.
     \item \textbf{Output generated by the LLM}: This is referred to as the assistant prompt. This is the text the model produces, in response to a user prompt. In our experiments, the assistant prompt corresponds to the assurance case review that the LLM generates for a given review criterion.
 \end{itemize}

 \subsubsection{Description of LLM User Prompt}
Our user prompt specifies the review criterion that the LLM should use to complete the review, as well as the name of the assurance case it needs to review. The user prompt also asks the LLM to display the results of its review, depending on the information the LLM specified in the system prompt. Figure \ref{fig:user-prompt} illustrates an example of a user prompt. That user prompt focuses on the safety case of a system called Baidu Apollo. 

 \smallskip
\noindent\fbox{%
    \parbox{\linewidth}{%
    \smallskip
    Using the Argument Comprehension criterion, review the safety case of the Baidu Apollo. Once finished display the results accordingly as to how the review should be conducted.
    }}
    \captionof{figure}{An example of User Prompt}
    \label{fig:user-prompt} 

 \subsubsection{Description of LLM System Prompt}
Our system prompts allow us to instruct the LLM on how it should conduct the review of the analyzed assurance case based on a specific criterion.  
We designed our system prompts to allow us to use LLMs as judges, i.e., in these prompts, we ask LLMs to rate the quality of the assurance case based on a given review criterion. The LLM specifies its rating by using a linear scale ranging from 1 to 5. 1 means the assurance case is perfect (i.e., it is totally correct in accordance with the review criterion), and no change is needed. 5 means the assurance case is totally flawed and therefore inadequate to support the certification of the system under scrutiny. The LLMs ratings allow us to gauge the perception LLMs have regarding the quality of the assurance cases manually created by assurance case developers.

\subsubsection{CoT design for System Prompts}
Some of our system prompts leverage CoT. To define the steps that the LLM must follow through the CoT, we adapted the methodology that \cite{yu2024fine} proposed in Sections 3.2.1 and 3.2.2 of their paper focusing on the code review task. More specifically, we relied on predicate-based logic to adapt the symbols and notations \cite{yu2024fine} proposed. This allowed us to create a set of predicates (i.e. mathematical notations) suitable for the assurance case review task. Table \ref{tab:notations} describes these predicates. We rely on these predicates to derive the predicate-based rules that we use to craft our CoT prompts templates.

We designed four CoT prompts templates, each focusing on one of the four review criteria. Tables \ref{tab:argument comprehension}, \ref{tab:well-formedness}, \ref{tab:expressive sufficiency}, and \ref{tab:argument criticism} report these templates. Using predicate-based rules, each of these templates specifies the main objectives of the review process done in accordance with a specific review criterion, then uses a CoT-like reasoning to break down the review of the analyzed assurance case into a set of intermediate steps consisting of reporting detected criterion-oriented issues in the assurance case and making suggestions to tackle them. If after finishing its review, the LLM concludes that the analyzed assurance case LLM is perfect (totally correct) and that it detected no review issues aligning with the given review criterion, the template instructs the LLM to provide in its review report a rationale explaining why it reached that conclusion.

\begin{table}[h!]
    \centering
    \caption{Mathematical Notations for the CoT Prompt Template}
    \begin{tabular}{p{6in}}
    \toprule
    Notations used in the CoT Template\\
    \midrule
     \textbf{Issue({E}, {D})}: This notation is used to describe any issues discovered by the LLM, E is used to identify the GSN label, D is used to identify the textual description of the GSN element. \\
     \textbf{Description({I(n)}, {E}, {ID})}: This notation is used to describe any issues that the LLMs were able to identify. I(n) is used to identify the issue number, E is used to identify the GSN label, ID is used to identify the short description of the issue. \\
     \textbf{Suggest({I(n)}, {E}, {Sg})}: This notation is used to identify the suggestions the LLMs give. I(n) is used to identify which issue number the LLMs are referring to, E is used to identify the GSN label, Sg is used to identify the solution the LLM recommends to solve the issue.\\
     \textbf{<({E(n)}, {IDEN(n)})>}: This notation is exclusively used on the Well-Formedness (Syntax) criterion, as it is used to identify identical GSN labels that were located by the LLMs. E(n) is used to identify identical GSN labels, IDEN(n) is used to identify the textual descriptions of the identical GSN labels.\\
     \textbf{Structural({E},{D})}: This notation is exclusively used on the Well-Formedness (Syntax) criterion, as it is used to identify the structural issues that were located by the LLMs. E is used to identify the GSN label, D is used to identify the textual description of the GSN element. \\
     \textbf{Defeaters({D(n)}, {Df}, {De})}: This notation is exclusively used on the Argument Criticism and Defeat criterion, as it is used to identify defeaters that are present in the assurance case. D(f) is used to identify the defeater number, Df is used to identify the defeater, De is used to identify the GSN label that is being challenged or defeated by the defeater.
     \\
    \bottomrule
    \end{tabular}
    \label{tab:notations}
\end{table}

\begin{table}[]
    \centering
    \caption{CoT Prompt Template for Argument Comprehension}
    \begin{tabular}{p{6in}}
    \toprule
    Argument Comprehension Prompt Template\\
    \midrule
    The Argument Comprehension criterion focuses on determining if the supporting evidence of the assurance case relates back to the main goal. The key points on reviewing the Argument Comprehension criterion are as follows:\\
    \\
    1. Determine if the supporting evidence of each GSN element is related to it and is able to satisfy the missing requirements without any question.\\
    2. Determining if each of the linked GSN elements is directly and indirectly linked to the main goal is crucial or important in satisfying the main goal.\\
    \\
    To review the Argument Comprehension criterion of an assurance case, there will be 3 subtasks that need to be followed and implemented accordingly, those 3 subtasks are as follows:\\
    \\
    \textbf{Identifying issues in the assurance case} - The first subtask is to identify any issues currently present in the assurance case. To denote any issues in the assurance case, the notation Issue(\${E: The current GSN label}, \${D: Textual description of the current GSN element}) will be used.\\
    \\
    \textbf{Description of issue} - Once an issue is identified, the next step is to describe the issue and give an explanation to why it is an issue. To describe the issue in the assurance case, the notation Description(\${I(n): The current issue number}, \${E: The current GSN label}, \${ID: Description of the issue}) will be used.\\
    \\
    \textbf{Suggestion of Improvement} - Once the issue is described, the next step is to suggest an improvement to help solve the issue. To suggest an improvement, the notation Suggest(\${I(n): The current issue number}, \${E: The current GSN label}, \${Sg: Possible solution}) will be used.\\
    \\
    If there are no issues in the assurance case, the 3 subtasks can be skipped and instead of going through those subtasks, the score will just be issued along with a description of why the assurance case is perfect.\\
    \bottomrule
    \end{tabular}
    \label{tab:argument comprehension}
\end{table}

\subsubsection{Description of LLM system prompt templates}

To prompt LLMs, we rely on system prompts that we derived from three system prompt templates. Each system prompt template focuses on each of the four selected review criteria. 
Tables \ref{tab:context} and \ref{tab:review_description} in the Appendix Section respectively report the contextual information and the review criterion descriptions we used to create our system prompt templates. The contextual information specifies the GSN notation and its semantics. This information allows the LLM to get a better understanding of GSN. Note that we used the same contextual information and review criterion descriptions across all three system prompt templates. We further describe the structure of these prompt templates in the remainder of this section.

Note that our prompt templates bear some similarities with the ones \cite{odu2024automatic} proposed to support the automatic creation of assurance cases from patterns. Still, unlike Odu et al., our templates align with the assurance case review task and leverage different predicate-based rules and CoT-like reasoning.

\paragraph{Zero-shot System Prompt template} The Zero-shot template as seen in Table \ref{tab:zero-shot} (see Appendices), focuses on specifying the contextual information allowing the LLM to understand GSN concepts, the structured prose version of the assurance case the LLM should review, the review criterion the LLM at hand should use to perform the review, and the description of the so-called review criterion.

\paragraph{Zero-shot with CoT System Prompt template}  The Zero-shot with CoT template as seen in Table \ref{tab:zero-shot-with-cot} (see Appendices), focuses on specifying the contextual information, the structured prose version of the assurance case the LLM should review, the review criterion, the description of the review criterion. It also provides a description of the CoT process specifying the various steps an LLM should use when performing the review based on a specific review criterion. Hence, we obtained the Zero-shot with CoT prompt template by combining the following two elements: 1) the Zero-Shot template that Table \ref{tab:zero-shot} depicts; and 2)  the CoT template created for a specific review criterion (i.e. the template reported in Table \ref{tab:argument comprehension}, Table \ref{tab:well-formedness}, Table \ref{tab:expressive sufficiency}, or Table \ref{tab:argument criticism}).

\paragraph{One-shot with CoT System Prompt template}  The One-shot with CoT template as seen in Table \ref{tab:one-shot-cot}, focuses on specifying the contextual information, the structured prose version of the assurance case the LLM should review, the review criterion, as well as the description of the review criterion. It also provides the description of the CoT reasoning embodying the various steps allowing to review an assurance case. The one-shot with CoT prompt template also specifies the review example i.e. one-shot example the LLM at hand should use to improve its reasoning capabilities. The review example consists of a given assurance case specified in the structured prose together with the outcome of the review manually performed on this assurance case. We therefore obtained the One-shot with CoT prompt template  by combining the following three elements: 1) the Zero-Shot template that Table \ref{tab:zero-shot} depicts; 2)  the CoT template created for a specific review criterion (i.e. the template reported in Table \ref{tab:argument comprehension}, Table \ref{tab:well-formedness}, Table \ref{tab:expressive sufficiency}, or Table \ref{tab:argument criticism}); and 3) a one-shot example obtained from a manual assurance case review. 

\begin{table}[]
    \centering
    \caption{One-Shot with CoT System Prompt Template}
    \begin{tabular}{p{7in}}
    \toprule
    One-Shot with CoT System Prompt Template\\
    \midrule
    You are an AI assistant who will assist in reviewing an assurance case represented using the Goal Structuring Notation (GSN). Your role as the assistant is to review the assurance case by scoring it using a linear scale ranging from 1 to 5, with 1 meaning the assurance case is totally correct and there is therefore no room for error and 5 meaning the assurance case is totally incorrect (i.e. full of errors). When the assurance case score is not 1, give a reasoning or give examples on how the assurance case can be further improved. More context about the assurance case will begin with the delimiter “@Context\_AC” and ends with the delimiter “@End\_Context\_AC” while the assurance case to be reviewed will begin with the delimiter “@Assurance\_Case” and end with the delimiter “@End\_Assurance\_Case”\\
    \textcolor{red}{@Context\_AC}\\
    More Context Information on the assurance case to be placed here\\
    \textcolor{red}{@End\_Context\_AC}\\
    \textcolor{red}{@Assurance\_Case}\\
    The Assurance Case to be reviewed should be specified here in the structured prose format complying with GSN\\
    \textcolor{red}{@End\_Assurance\_Case}\\
    We have also defined which review criterion will be used for the review of an assurance case. This criterion will solely be used as a basis on how the assurance case should be reviewed and scored. The review criterion to be used will begin with the delimiter  “@Review\_Criterion” and end with the delimiter “@End\_Review\_Criterion” while the description of the review criterion will start with the delimiter “@Review\_Criterion\_Description” and end with the delimiter “@End\_Review\_Criterion\_Description” with the chain of thought process on how the review should be done incrementally would begin with the delimiter “@Chain\_Of\_Thought” and end with the delimiter “@End\_Chain\_Of\_Thought”.
     An example under the form of an assurance case together with its manual review is also included. This example assurance case starts with the delimiter “@Assurance\_Case\_Review\_Example” and ends with “@End\_Assurance\_Case\_Review\_Example” while its manual review starts with the delimiter “@Example\_Of\_Review\_Done\_On\_The\_Example\_Assurance\_Case” and ends with “@End\_Example\_Of\_Review\_Done\_On\_The\_Example\_Assurance\_Case”. These two will help become familiar with how an assurance case is reviewed.\\
    \textcolor{red}{@Review\_Criterion}\\
    Name of the review criterion to be specified here\\
    \textcolor{red}{@End\_Review\_Criterion}\\
    \textcolor{red}{@Review\_Criterion\_Description}\\
    Description of the review criterion to be placed here\\
    \textcolor{red}{@End\_Review\_Criterion\_Description}\\
    \textcolor{red}{@Chain\_Of\_Thought}\\
    Chain of thought text to be added here\\
    \textcolor{red}{@End\_Chain\_Of\_Thought}\\
    \textcolor{red}{@Assurance\_Case\_Review\_Example}\\
    Example of manually reviewed assurance case to be added here \\
    \textcolor{red}{@End\_Assurance\_Case\_Review\_Example}\\
    \textcolor{red}{@Example\_Of\_Review\_Done\_On\_The\_Example\_Assurance\_Case}\\
    Outcomes of the manual review done on the example assurance case to be added here\\
    \textcolor{red}{@End\_Example\_Of\_Review\_Done\_On\_The\_Example\_Assurance\_Case}\\
    \bottomrule
    \end{tabular}
    \label{tab:one-shot-cot}
\end{table}





\section{Experimental Setup}
\label{sec:experimental_setup}

\subsection{Research Questions}

Our experiments aim at investigating the following research questions (RQs):
\newline
\textbf{(RQ1): When completing the review process, how do the LLMs perceive the quality of assurance cases manually generated by assurance case developers?}
\newline
\textbf{(RQ2): Are LLMs effective at automatically reviewing existing assurance cases that were manually created by experts?}
\newline
\textbf{(RQ3): Which prompting strategy yields the best LLM-based assurance case reviews?}
\newline
\textbf{(RQ4): Which LLM generates the best assurance case reviews?}
\newline
\textbf{(RQ5): What are the typical review capabilities and issues that the LLMs at hand exhibit?}

\subsection{Dataset Description}
We followed the process explained in Phase I of our approach (see Section \ref{subsec:data-collection}
) to collect assurance cases in the literature. Our experiments, therefore, focus on five manually created assurance cases represented using GSN: four assurance cases that we attempt to automatically review using LLMs, as well as one additional assurance case that we use as an example in our One-Shot with CoT prompts. Our experiments therefore, focus on the following assurance cases that Table \ref{table:dataset_overview} describes: 1) the safety case of a Lane Management System (LMS) by \cite{di2019querying}; 2) the safety case of the trajectory prediction component of Baidu Apollo that we adapted from \cite{odu2025llm}; 3) the safety case of GPCA (Generic Patient-Controlled Analgesia) by \cite{lin2017support}; 4) the security case of an Instant messaging (IM) server software by \cite{xu2017layered}; and 5) the safety case of a Level 4 Automated Driving (i.e. high driving automation) system by \cite{inbook}.

\begin{table} [!h]
\centering
\caption {Dataset Overview - adapted from \cite{odu2024automatic} }
\label{table:dataset_overview}

 \begin{tabular}{|p{5.0cm} |p{2.5cm} |p{2.5cm} |p{3.0cm}  |p{3.0cm}  |}

\hline
\multirow{2}{*}{\textbf{System}} & \multirow{2}{*}{\textbf{Domain}}& \multicolumn{3}{c|}{\textbf{Assurance Case (ACs)}} \\
 
\cline{3-5}
 & & \textit{\textbf{Decorators}}& \textit{\textbf{Elements}} & \textit{\textbf{Relationships}} \\
\hline
Baidu Apollo & Automotive & 0 & 38 & 41 \\
\hline
GPCA & Medical & 6 & 27 & 26  \\
\hline
IM server SOFTWARE & Computing & 0 & 24 & 23 \\
\hline
LMS & Automotive & 0 & 76 & 77 \\
\hline
Level 4 Automated Driving System 
& Automotive & 0 & 38 & 37\\
\hline

\end{tabular}

\end{table}

As we specified in Phase II of our approach, we need to convert each GSN-compliant assurance case in its structured prose format to ease its ingestion by the LLM at hand. Still, the structured prose version of most of these assurance cases is publicly available in literature (e.g., \citep{sivakumar2024prompting, odu2025llm}). So, we only needed to manually generate the structured prose version of the following assurance case: Level 4 Automated Driving System, IM Software, and GPCA.

Note that some GSN elements of the GPCA safety case and of the IM Server Software security case have duplicated identifiers, whereas GSN  fosters the uniqueness of identifiers, which is one of its core rules. 
Furthermore, due to space restrictions, the authors of the paper describing the IM Server Software security case depicted some of its goals without supporting arguments or evidence. These are examples of structural errors that the review based on the Well-Formedness criterion should flag.

\subsection{LLM Description and Settings}

We rely on four state-of-the-art Large Language Models to automate our review approach: \textbf{Gemini 2.0 Flash}, \textbf{GPT-4o}, \textbf{GPT-4.1}, and \textbf{DeepSeek-R1}. We used each of them with its default settings.
 Table \ref{tab:LLM overview} reports some characteristics of these LLMs. These characteristics include which organization developed these LLMs, their release date, and their cut-off date. 
      
\begin{table}[h!]
    \centering
    \caption{Overview of the LLMs}
    \begin{tabular}{p{2in}p{1.2in}p{1.2in}p{1.2in}}
    \hline
    LLM& Organization & Release Date & Cut-off Date\\
    \hline
    \textbf{Gemini 2.0 Flash} &  Google & February 5, 2025 & June 2024\\
    \textbf{GPT-4o} & OpenAI & May 2024 & October 2023\\
    \textbf{GPT-4.1} & OpenAI & May 14, 2025 & June 2024\\
    \textbf{DeepSeek-R1} & DeepSeek-AI & January 20, 2025 & January 2025\\
    \hline
    \end{tabular}
    \label{tab:LLM overview}
\end{table}

\subsection{Experiments Description}
To investigate our research questions, we performed the following experiments by relying on the prompts we designed in Phase IV of our approach (see Section \ref{subsec:prompt-engineering}):

\begin{itemize}
    \item \textbf{Experiment 1 (Zero-shot)}: To carry out this experiment, we used our Zero-shot template to derive the zero-shot prompts we used to run each LLM. We fed each of the four analyzed assurance cases in the structured prose format as input to the LLM. Each zero-shot prompt focuses on the review of one of the four assurance cases based on one of the four review criteria. We therefore derived sixteen zero-prompts from our Zero-shot template. 
\item \textbf{Experiment 2 (Zero-shot + CoT)}: 
To perform this experiment, we used our Zero-shot + CoT template to run each LLM, using each of the four analyzed assurance cases in structured prose as input. For this purpose, we derived sixteen zero-shot + COT prompts from our Zero-shot+COT template. 
\item \textbf{Experiment 3 (One-shot + CoT)}: 
To carry out this experiment, we used our One-shot + CoT template to run each LLM, using as input each of the four analyzed assurance cases in the structured prose format together with an example. We therefore derived sixteen one-shot + COT prompts from our one-shot + COT template.  
\end{itemize}

In accordance with the literature (e.g., \citep{chen2023use, sivakumar2024prompting, odu2024automatic}), we ran each experiment five times to account for the non-determinism of LLMs. When carrying out each experiment, we therefore relied on sixteen prompts to sequentially query each of the four LLMs five times. Each experiment therefore yielded a total of 320 LLM-generated assurance case review outputs.

\subsection{Assessment Metrics}
Due to proprietary and security concerns, there is no publicly available ground truth we can use as a reference when assessing our experimental results. This prevents us from using common metrics such as Precision, Recall, F-Measure, and BLEU score. Hence, following the guidelines for empirical studies in Software Engineering involving LLMs\cite{baltes2025evaluation}, we rely on human validation to assess LLM outputs (i.e. LLM-generated reviews). Thus, to assess the results the LLMs generate during our experiments, we rely on the following metrics that the well-grounded work of \cite{viger2024ai} proposed.

\begin{itemize}

    \item \textbf{Informativeness}: Informativeness is the ability of the LLMs to be able to describe or explain concretely the answers it has generated (What, Where or Why responses?) \citep{viger2024ai}. In the context of reviewing assurance cases, the informativeness will be rated on how the LLMs are able to explain or describe how their review process is done. Along with being able to properly describe their review process, the LLMs should also be able to properly describe individually the issues the LLMs have managed to detect when reviewing the assurance cases. The informativeness is also rated on how properly they can explain the possible improvements that can be done on those "flawed" assurance cases. We assess the Informativeness of the response generated by the LLMs based on a linear scale ranging from 1 to 5. We interpret that scale as follows \textbf{1=Totally Informative; 2=Mostly Informative; 3=Moderately Informative; 4=Slightly Informative; 5=Uninformative}. 

     \item \textbf{Coherence}: Coherence is the trait of an LLM on how frequently its answers generated contain hallucinations and the severity of those hallucinations. Hallucinations are answers generated by LLM that are factually incorrect or nonsensical information, and having those generated answers be stated as factual by the LLM \citep{10569238}. In the context of assurance case review, the coherence would be rated based on how often or how impactful those hallucinations that appear in the assurance case review can be. This might either appear in the form of as a recommendation of improvement on a non-existent or non-related scenario in an assurance case. As with the above metric, we assess the Coherence of the response the LLMs yield by relying on a linear scale ranging from 1 to 5. On this scale, 1 means \textbf{Totally Coherent}, while 5 means \textbf{Incoherent}. 

      \item \textbf{Usefulness}: Usefulness is the ability of an LLM to generate an answer that is considered helpful or insightful for the argument developer. In our context, usefulness would be determined by the extent to how useful or insightful the generated review of an LLM can be whether as a stand-alone reviewer or an assistant to a person manually reviewing an assurance case. As with the two other metrics, we assess the Usefulness of the review generated by each LLM based on a linear scale ranging from 1 to 5. On this scale, 1 means \textbf{Totally Useful}, while 5 means \textbf{Not Useful}. 
\end{itemize}

It is well-known that relying on human judgment is the best way to evaluate LLMs \citep{gu2024survey}. So, as in \citep{viger2024ai},  two assessors having at least five years of experience in the software engineering field and having a good expertise in system assurance were involved in the assessment of the aforementioned metrics. Howewer, collecting human judgment is usually tedious, costly, and time-consuming \citep{gu2024survey,ouyang2022training, szymanski2025limitations, zheng2023judging}. Hence, the first assessor (i.e., senior co-author) and the second assessor (i.e. another co-author) used a linear scale ranging from 1 to 5 to manually and separately review a random sample of the LLM-generated reviews the three experiments yield. The two assessors then met multiple times to establish clear, consistent, and rigorous guidelines on how to manually rate the LLM-generated results. In accordance with these guidelines, the second assessor then relied on a linear scale ranging from 1 to 5 to manually and independently assign a value to each of the three metrics. The second assessor therefore assessed all the LLM-generated reviews that the three experiments yield by manually specifying the ratings associated with each metric at hand. 
The outcomes of these assessments are presented in Section \ref{sec: results}. 

\section{Results and Analysis} \label{sec: results}

For each of the LLM, each analyzed assurance case, each prompting strategy, and each of the review criteria,  the three experiments collectively generated a total of 960 LLM-based assurance case reviews (i.e. review reports) over five runs. In this section, we discuss the results the second assessor obtained when manually assessing the LLM-generated results  that the three experiments yield.

\subsection{(RQ1) LLM perception Regarding the Quality of the Assurance Cases manually created by Assurance Case Developers}
\label{subsec:RQ1-results}

Table \ref{tab:RQ1_llm_ratings} reports the average of the LLM-based reviews for each review criterion over five runs. As this Table illustrates, all LLMs consistently assigned a rating of 2 or 3 for each review criterion, the majority of ratings being closer to 3. Overall, they therefore consensually judged the analyzed assurance cases as moderately correct, indicating a need for further refinement. Furthermore, out of the four LLMs at hand, DeepSeek-R1 is the harshest of them since all its ratings are usually the highest. Contrariwise, GPT-4.1 and GPT-4o are the least severe of all four LLMs under analysis.

\begin{table}[h!]
\centering
\renewcommand{\arraystretch}{1.1}
\setlength{\tabcolsep}{8pt}
\begin{tabular}{|l|l|c|c|c|c|}
\hline
\textbf{LLM} & \textbf{Strategy} & \multicolumn{4}{c|}{\textbf{LLM Ratings}} \\
\cline{3-6}
 & & \textbf{AV A. C.} & \textbf{AV W-F.} & \textbf{AV Exp. Suf.} & \textbf{AV Arg. Cri.}\\
\hline
\textbf{GPT-4o} & ZS & 2.65 & 2.6 & 2.9 & 3 \\
& ZS with CoT & 3.25 & 2.85 & 3.15 & 3.075 \\
& OS with CoT & 2.9 & 2.6 & 3.1 & 3.4 \\
\hline
\hline
\textbf{GPT 4.1} & ZS & 2.5 & 2.7 & 2.6 & 2.65 \\
& ZS with CoT & 3.05 & 3.45 & 3.3 & 3.2 \\
& OS with CoT & 2.95 & 3.15 & 3.2 & 3.075 \\
\hline
\hline
\textbf{DeepSeek-R1} & ZS & 3.5 & 3.45 & 3.75 & 3.65 \\
& ZS with CoT & 3.6 & 3.05 & 3.65 & 3.375 \\
& OS with CoT & 3.75 & 3.45 & 3.6 & 3.375 \\
\hline
\hline
\textbf{Gemini 2.0 Flash} & ZS & 3.2 & 2.95 & 3 & 3.2 \\
& ZS with CoT & 3.25 & 3.35 & 3.65 & 3.3 \\
& OS with CoT & 2.9 & 3.15 & 3 & 2.8 \\
\hline
\end{tabular}
\caption{Average of the LLM-Based Review Ratings Across Five Runs and Across the Four Analyzed Assurance Cases (ZS = Zero-shot; OS = one-shot)}
\label{tab:RQ1_llm_ratings}
\end{table}

To have a finer-grained understanding of the extent to which LLMs reached a consensus when judging the quality of the analyzed assurance cases, we quantitatively assessed the inter-model agreement. For this purpose, we relied on a statistical measure called Fleiss' kappa (\textit{K}) \citep{fleiss1971measuring, warrens2010inequalities}. The values of the Fleiss kappa vary between -1 (total disagreement) and 1 (total agreement).  We automatically computed the values of this measure using two Python libraries: \textbf{statsmodels.stats.inter\_rater.fleiss\_kappa}\footnote{Website for Fleiss Kappa library: {\scriptsize \url{https://www.statsmodels.org/dev/generated/statsmodels.stats.inter_rater.fleiss_kappa.html}}} and \textbf{statsmodels.stats.inter\_rater.aggregate\_raters}\footnote{Website for Aggregate Raters: {\scriptsize \url{https://www.statsmodels.org/dev/generated/statsmodels.stats.inter_rater.aggregate_raters}}}. 
As Table \ref{tab:RQ1_fleiss_kappa_ratings} shows,  the values of the Fleiss' kappa are negative for most of the review criteria and most of the prompting strategies. So, LLMs therefore tend to disagree in their individual ratings, and these ratings are not always consistent. This means that, while LLMs seem to yield quite similar global ratings when it comes to the assurance cases quality (see Table  \ref{tab:RQ1_llm_ratings}), they seem to have quite different opinions and interpretations of what the quality of assurance cases entails when it comes to the four review criteria at hand.

\begin{table}[h!]
\centering
\renewcommand{\arraystretch}{1.1}
\setlength{\tabcolsep}{8pt}
\begin{tabular}{|l|l|c|}
\hline
\textbf{Review Criterion} & \textbf{Strategy} & \textbf{Fleiss Kappa} \\
\hline
\textbf{Argument Comprehension} & ZS & -0.0928 \\
& ZS with CoT & 0.0342  \\
& OS with CoT & -0.0433  \\
\hline
\hline
\textbf{Well-Formedness(Syntax)} & ZS & 0.0560  \\
& ZS with CoT & -0.0214  \\
& OS with CoT & -0.0751  \\
\hline
\hline
\textbf{Expressive Sufficiency} & ZS & -0.1616 \\
& ZS with CoT & -0.0218 \\
& OS with CoT & -0.0092 \\
\hline
\hline
\textbf{Argument Criticism and Defeat} & ZS & -0.1542 \\
& ZS with CoT & -0.0009 \\
& OS with CoT & 0.1453 \\
\hline
\end{tabular}
\caption{Fleiss Kappa Ratings of the Four LLMs on each Prompting Strategy (ZS = Zero-shot; OS = One-shot)}
\label{tab:RQ1_fleiss_kappa_ratings}
\end{table}

\smallskip
\noindent\fbox{%
    \parbox{\linewidth}{%
    \smallskip
    \textbf{Key Takeaways (RQ1):} The analysis of the results shows that even though human experts manually crafted the assurance cases under scrutiny, LLMs still rated their quality as moderate based on the four review criteria at hand.
    }}

\subsection{(RQ2): Effectiveness of LLMs in Automatic Assurance Case Review}

\subsubsection{Informativeness}

Table \ref{tab:RQ2_informativeness} reports the averages of the informativeness ratings specified by the second assessor for each analyzed assurance case, for each LLM at hand, and for each review criterion over the five runs of each experiment. 

 Table \ref{tab:RQ2_informativeness} shows that with all the CoT prompting strategies, all the LLMs can achieve an informativeness score close to one (i.e., Totally Informative) for most of the four review criteria, with DeepSeek-R1  and GPT 4.1 being the most informative models overall. More specifically, these LLMs consistently yield Informativeness scores close to 1 for the following review criteria: Argument Comprehension, Expressive Sufficiency, and Argument Criticism and Defeat. However, these LLMs usually yield higher (i.e. poorer) Informativeness scores for the following review criterion: Well-Formedness (Syntax). Hence, for these review criteria, human experts still need to refine the reviews that LLMs generate.

Interestingly, when relying on the Zero-shot prompting strategy, all four LLMs, and particularly GPT 4.1,  were the least informative when providing suggestions to improve the review issues they detected. The rationale is that the zero-shot prompting strategy lacks a structured way to instruct LLMs on reporting detected review issues as well as suggestions to fix them, leading them to generate generic and therefore significantly less informative review issue descriptions and fix suggestions. 

\begin{table}[h!]
\centering
\renewcommand{\arraystretch}{1.1}
\setlength{\tabcolsep}{8pt}
\begin{tabular}{|l|l|c|c|c|c|}
\hline
\textbf{LLM} & \textbf{Strategy} & \multicolumn{4}{c|}{\textbf{Informativeness }} \\
\cline{3-6}
 & & \textbf{AV A. C.} & \textbf{AV W-F.} & \textbf{AV Exp. Suf.} & \textbf{AV Arg. Cri.}\\
\hline
\textbf{GPT-4o} & ZS & 3.65 & 3.2 & 3.4 & 3.65 \\
& ZS with CoT & 1.95 & 1.65 & 1.35 & 1.7 \\
& OS with CoT & 1.5 & 2.05 & 1.2 & 1.3 \\
\hline
\hline
\textbf{GPT 4.1} & ZS & 3.6 & 3.05 & 3.1 & 2.95 \\
& ZS with CoT & 1.6 & 1.3 & 1.4 & 1.65 \\
& OS with CoT & 1.25 & 1.5 & 1.35 & 1.2 \\
\hline
\hline
\textbf{DeepSeek-R1} & ZS & 3.35 & 2.65 & 2.85 & 3 \\
& ZS with CoT & 1.4 & 1.15 & 1.4 & 1.25 \\
& OS with CoT & 1.3 & 1.3 & 1.25 & 1.1 \\
\hline
\hline
\textbf{Gemini 2.0 Flash} & Zero-shot & 3.45 & 3.25 & 3 & 2.85 \\
& ZS with CoT & 1.8 & 1.6 & 1.7 & 1.55 \\
& OS with CoT & 1.45 & 1.65 & 1.7 & 1.2 \\
\hline
\end{tabular}
\caption{Average (Across Five Runs and Across the Four Analyzed Assurance Cases) of the Informativeness Results Manually Specified by the Second Assessor (ZS = Zero-shot; OS = one-shot)}
\label{tab:RQ2_informativeness}
\end{table}

\begin{tcolorbox}[rounded corners, colback=gray!15!white, colframe=gray!70!white, title=Sample of review demonstrating the informativeness of DeepSeek-R1]
Issue(G3, "Overinfusion" is mitigated) 

Description(I1, G3, Incomplete evidence: No supporting evidence or argument is provided to show how overinfusion is mitigated, such as test results or analyses.)

Suggest(I1, G3, Add concrete evidence such as infusion pump test results, hazard analysis reports, or clinical validation data to demonstrate overinfusion mitigation.)

\end{tcolorbox}
When examining the GPCA safety case provided in the Appendix, we can see that it comprises a goal labeled G3. This goal states the following: \textit{"Overinfusion" is mitigated}. However, this goal is undeveloped i.e. lack supporting GSN elements (e.g., evidence). As a result, its argumentation structure is difficult to understand. The textbox above shows that DeepSeek-R1 is able to clearly spot this issue, describe it, and make a suggestion to fix this issue. This textbox presents an excerpt from the review that DeepSeek-R1 generated for the GPCA safety case when using the One-Shot with CoT prompt, and focusing on the Argument Comprehension criterion. This demonstrates that DeepSeek-R1 can provide highly informative feedback when detecting and describing issues within an assurance case.



\subsubsection{Coherence}

Table \ref{tab:RQ2_coherence} shows the averages of the coherence ratings specified by the second assessor for each analyzed assurance case, for each LLM at hand, and for each review criterion over the five runs of each experiment. 

Interestingly, when relying on the two CoT prompting strategies, all four LLMs achieve relatively good coherence scores that are usually close to two (i.e. Mostly Coherent) for three of the models (i.e. DeepSeek-R1, GPT 4.1, and Gemini 2.0 Flash) and close to three (i.e. Moderately Coherent) for GPT-4o. This indicates that, when provided with sufficient instructions thanks to the use of the CoT prompting strategies, the four LLMs are able to create reviews that are often factually correct. Still, regardless of the prompting strategy at hand, all LLMs sometimes hallucinate. The hallucinations the LLMs generate are quite diverse. In some hallucinations, LLMs generate information that is untrue or is not present, such as referencing some GSN elements that are not present in the analyzed assurance case. Other hallucinations include not being able to specify all the GSN elements that are identical, even though they are instructed to list out all duplicated GSN elements. Another instance of hallucinations is when the LLMs do not flag an issue on GSN elements that do not follow the naming conventions promoted by the GSN standard.

\begin{tcolorbox}[rounded corners, colback=gray!15!white, colframe=gray!70!white, title=Example of hallucination generated by Gemini 2.0 Flash when reviewing the safety case of GPCA]

Issue(G3, “Overinfusion” is mitigated)

Issue(G3, “Underinfusion” is mitigated)

Description(I1, G3, Incomplete evidence: The two instances of G3 lack supporting evidence or further decomposition to demonstrate how overinfusion and underinfusion are mitigated. Without this evidence, it's impossible to assess the argument's comprehension regarding hazard mitigation.)

Suggest(I1, G3, Provide Evidence: Add solution nodes (Sn) with references to tests, analyses, or design features that demonstrate how overinfusion and underinfusion are mitigated. Alternatively, decompose G3 into sub-goals that address specific mitigation measures.)
\end{tcolorbox}
The textbox above showcases a sample of the review  Gemini 2.0 Flash generated when analyzing the safety case of GPCA using the One-Shot with CoT prompt. This review focuses on the Argument Comprehension criterion. As shown in this review, Gemini 2.0 Flash indicated that there are issues with two GSN elements respectively called G3. The issue being that both elements are undeveloped and therefore lacking supporting evidence. But, when looking at the safety case of GPCA (see the Appendix Section), we can clearly see that G3 is a duplicated GSN element: the first instance of G3 is indeed undeveloped, but the second instance of G3 is not undeveloped because it is clearly supported by other GSN elements, including solutions (i.e. evidence). So, the duplication of G3 in the GPCA safety case clearly confused Gemini 2.0 Flash, driving it to generate non-sensical information about the second instance of G3. 
This example therefore shows that while Gemini 2.0 Flash is able to perform complex tasks such as the review of assurance cases, this model is not immune to hallucinations, a common flaw of LLMs.


\begin{table}[h!]
\centering
\renewcommand{\arraystretch}{1.1}
\setlength{\tabcolsep}{8pt}
\begin{tabular}{|l|l|c|c|c|c|}
\hline
\textbf{LLM} & \textbf{Strategy} & \multicolumn{4}{c|}{\textbf{Coherence }} \\
\cline{3-6}
 & & \textbf{AV A. C.} & \textbf{AV W-F.} & \textbf{AV Exp. Suf.} & \textbf{AV Arg. Cri.}\\
\hline
\textbf{GPT-4o} & ZS & 3.15 & 2.55 & 2.85 & 3 \\
& ZS with CoT & 2.2 & 2.45 & 2.2 & 2.6 \\
& OS with CoT & 2.95 & 2.85 & 2.8 & 2.7 \\
\hline
\hline
\textbf{GPT 4.1} & ZS & 2.3 & 2 & 2.25 & 2.1 \\
& ZS with CoT & 2.6 & 1.75 & 2.35 & 2.2 \\
& ZS with CoT & 2.15 & 1.75 & 2.45 & 2.3 \\
\hline
\hline
\textbf{DeepSeek-R1} & ZS & 2.6 & 2.3 & 2.25 & 2.55 \\
& ZS with CoT & 2.25 & 1.95 & 2.2 & 2.05 \\
& OS with CoT & 1.95 & 1.5 & 2.1 & 2.15 \\
\hline
\hline
\textbf{Gemini 2.0 Flash} & ZS & 2.35 & 2.65 & 2.5 & 3.6 \\
& ZS with CoT & 2.5 & 2.35 & 2.4 & 2.7 \\
& OS with CoT & 2.55 & 2.6 & 2.4 & 2.35 \\
\hline
\end{tabular}
\caption{Average  (Across Five Runs and Across the Four Analyzed Assurance Cases) of the Coherence Results Manually Specified by the Second Assessor (ZS = Zero-shot; OS = one-shot)}
\label{tab:RQ2_coherence}
\end{table}

\subsubsection{Usefulness}

Table \ref{tab:RQ2_usefulness} reports the averages of the Usefulness ratings for each analyzed assurance case, for each LLM at hand, and for each review criterion over the five runs of each experiment. 

Interestingly, as in the case of the Informativeness, the Usefulness of suggestions provided by the LLMs decreases when using the Zero-shot prompting strategy. This is probably because this strategy does not leverage any specific mathematical notation or examples as to how a suggestion should be formulated to solve the issues that the LLMs have detected. The use of this strategy therefore drive LLMs to generate fix suggestions that are often vague and generic. Although the LLMs perform badly on the zero-shot prompting strategy, DeepSeek-R1, as observed, tended to perform better than the other three LLMs. This is because, despite the general suggestions still being generic and vague, DeepSeek-R1 still manages to provide more concrete suggestions on how to improve analyzed assurance cases. 

Another interesting observation is that, while the Usefulness of the suggestions generated by the LLMs seems to improve when they leverage the CoT prompts, these improvements are more noticeable on the Well-Formedness (Syntax) criterion, except for Gemini 2.0 Flash. The rationale is that, unlike other LLMs, Gemini 2.0 Flash tends to omit some suggestions in its review. For instance,  even when it identifies duplicated GSN elements in an analyzed assurance case, it sometimes fails to mention them in its suggestions for improvement. Furthermore, LLMs struggle to be useful when reviewing assurance cases based on the Expressive Sufficiency criterion. This is because LLMs tend to flag issues focusing on external artifacts or documentations accompanying the manual assurance cases and mentioned in some of their GSN elements. Since these documents were not publicly available and therefore not supplied as inputs to LLMs, these LLMs lack the necessary context to reason about them and incorrectly flag them as problematic. Flagging such irrelevant issues is not useful for professionals who already have contextual information on these valid external artifacts and documentations accompanying the assurance case at hand.

\begin{tcolorbox}[rounded corners, colback=gray!15!white, colframe=gray!70!white, title=Example of defeater DeepSeek-R1 generated for the Baidu apollo safety case]
Issue(Sn6.2, Operational testing logs are shared between G6.5 and G6.6, which may not adequately support both design and operational aspects if the evidence is not comprehensive.)

Defeater(D4, If the operational testing logs (Sn6.2) do not comprehensively cover both design and operational scenarios, they may not sufficiently support G6.5 and G6.6, leading to a defeat in the deployment argument.)
\end{tcolorbox}
The textbox above shows a sample of the review  DeepSeek-R1 generated when analyzing the safety case of Baidu Apollo using the One-Shot with CoT prompt. This review focuses on the Argument Criticism and Defeat criterion. As shown in the textbox, DeepSeek-R1 can generate a rebuttal defeater that challenges the validity of a deployment argument by undermining its supporting evidence i.e., the solution called Sn6.2. This defeater indicates that if Sn6.2 does not extensively cover how the design and operation of the trajectory prediction component of Baidu Apollo works during its deployment, this solution may be inadequate to support its parent goals (i.e. G6.5 and G6.6). 
However, when generating this review, DeepSeek-R1 did not strictly follow the formalism we instructed it to use through the CoT prompt focusing on the Argument Criticism and Defeat (see Table \ref{tab:argument criticism}).



\begin{tcolorbox}[rounded corners, colback=gray!15!white, colframe=gray!70!white, title=Example of successful duplicate identification]
**Step 3: Identifying identical GSN labels**

- <(G3, “Overinfusion” is mitigated, G3, “Underinfusion” is mitigated)>

- <(G4, “Underinfusion” is mitigated under “Programmed flow rate too low”, G4, “Underinfusion” is mitigated under “Flow rate does not match programmed rate”)>

- <(G5, “SR1.2” is appropriate for “Flow rate does not match programmed rate”, G5, “SR6.1.4”  is appropriate for “Programmed flow rate too low”)>

- <(G6, “period is 15 mins” is appropriate for “SR1.2” over properties, G6, “flow rate sensor is equipped, “ is appropriate for “SR1.2”)>

- <(G7, “FDA standard” is appropriate and trustworthy, G7, “period is 15 mins” definition is is sufficient)>
\end{tcolorbox}
The textbox presents an excerpt of the review of GPT 4.1 generated when analyzing the GPCA safety case using the One-Shot with CoT prompt. This review focuses on the Well-Formedness (Syntax) criterion. 
As shown in this excerpt, GPT 4.1 is able to successfully identify all identical GSN elements (i.e. all the duplicated GSN elements) when analyzing the GPCA safety case. These elements are G3, G4, G5, G6, and G7. This shows that GPT 4.1 is often capable of identifying such structural issues when reviewing assurance cases. 


\subsubsection{Global Analysis of RQ2 Results}
\label{subsubsec:analaysys-rq1}

The manual ratings for the LLM-generated reviews indicate that the quality of the LLM-generated reviews usually varies between 2 and 3 in the case of CoT prompts, except for Informativeness, and between 3 and 4 when relying on zero-shot prompts. This indicates that, LLM-generated reviews are usually coherent, useful, and informative, but there is still room for improvement. For instance, all four LLMs, when presented with assurance cases that are incomplete (i.e, marked with decorators such as \textit{undeveloped}), are not able to consistently identify or reason about the corresponding GSN elements. More specifically, they sometimes struggle to identify any of them. This is the case for the assurance cases of both the GPCA system and the IM Server system. Of course, this is an extreme case because any assurance case that is being evaluated is expected to be complete, but to err is human, and extreme cases should therefore not be ruled out. This suggests we can not yet totally rely on automated LLM-generated reviews.

Interestingly, out of the four review criteria, LLMs usually yield the best ratings when using the Well-Formedness (Syntax) criterion to assess assurance cases. Accordingly, while LLMs are not able to consistently identify all duplicated GSN elements in an assurance case, they are still able to consistently identify at least half of them. However, LLMs struggle the most with the Expressive Sufficiency criterion because, as mentioned above, they lack knowledge about the various external artifacts and documents accompanying assurance cases. As a consequence, human oversight is necessary to complete and/or refine LLM-generated reviews.

Additionally, although the manual ratings for the LLM-generated reviews are promising, the metrics used to assess them may not fully capture all the issues we spotted when analyzing LLM-generated reviews. These issues include failure to sometimes use the specified mathematical notation when relying on the CoT prompts. Due to this, our assessment metrics may not fully capture all the review insights needed to fully grasp the strengths and weaknesses of LLMs in the context of assurance case review. The qualitative analysis we perform in Section \ref{subsec:taxonomy} addresses this gap. Still, given this limitation, we believe it is crucial for future research to also focus on developing more suitable metrics to support quantitative assessment. This observation echoes the recent recommendation by \cite{diemert2025one}  that invited the assurance case community to establish methodological standards that better support the evaluation of results in the emerging field of LLM-based system assurance.

\begin{table}[h!]
\centering
\renewcommand{\arraystretch}{1.1}
\setlength{\tabcolsep}{8pt}
\begin{tabular}{|l|l|c|c|c|c|}
\hline
\textbf{LLM} & \textbf{Strategy} & \multicolumn{4}{c|}{\textbf{Usefulness }} \\
\cline{3-6}
 & & \textbf{AV A. C.} & \textbf{AV W-F.} & \textbf{AV Exp. Suf.} & \textbf{AV Arg. Cri.}\\
\hline
\textbf{GPT-4o} & ZS & 4.65 & 3.85 & 4.55 & 4.55 \\
& ZS with CoT & 3.3 & 2.15 & 2.9 & 3.45 \\
& OS with CoT & 3.15 & 2.55 & 3.45 & 3.35 \\
\hline
\hline
\textbf{GPT 4.1} & ZS & 4.15 & 3.6 & 3.85 & 3.85 \\
& ZS with CoT & 3 & 1.85 & 3.4 & 3.2 \\
& OS with CoT & 3.05 & 2.35 & 3.75 & 3.15 \\
\hline
\hline
\textbf{DeepSeek-R1} & ZS & 3.5 & 3.15 & 3.4 & 3.4 \\
& ZS with CoT & 3.15 & 1.55 & 3.15 & 2.7 \\
& OS with CoT & 2.8 & 1.9 & 3 & 3 \\
\hline
\hline
\textbf{Gemini 2.0 Flash} & ZS & 3.85 & 3.9 & 4.25 & 4.15 \\
& ZS with CoT & 3.7 & 3.2 & 3.95 & 3.55 \\
& OS with CoT & 3.5 & 3.5 & 3.9 & 3.55 \\
\hline
\end{tabular}
\caption{Average  (Across Five Runs and Across the Four Analyzed Assurance Cases) of the Usefulness Results Manually Specified by the Second Assessor (ZS = Zero-shot; OS = one-shot)}
\label{tab:RQ2_usefulness}
\end{table}

\smallskip
\noindent\fbox{%
    \parbox{\linewidth}{%
    \smallskip
    \textbf{Key Takeaways (RQ2):}
    The analyses of the results we obtained for RQ2 suggest that, although LLMs are able to automatically generate quite informative, useful and informative reviews, humans are still needed to refine the LLM-generated reviews to improve their quality.
    }}

\subsection{(RQ3):  Prompting Strategy Yielding the Best LLM-based Assurance Case Reviews}

 We rely on Table \ref{tab:llm_comparison} 
 to answer RQ3. This Table
 reports for each prompting strategy, for each analyzed assurance case, for each assessment metric, and for each review criterion,  the averages of the ratings the second assessor provided when manually assessing LLM-based review results. As observed in Table \ref{tab:llm_comparison}, most LLMs, especially GPT-4o and Gemini 2.0 Flash, performed the worst when using the zero-shot prompting strategy. On the contrary, they performed the best when relying on the One-shot with CoT prompting strategy, with the sole exception of GPT-4o, where Zero-shot with CoT performed slightly better. So, when presented with a clear review example (i.e. a one-shot example), and when given detailed step-by-step instructions on how to perform the assurance case review task, LLMs reasoning capabilities are decoupled and their results are significantly improved.

\smallskip
\noindent\fbox{%
    \parbox{\linewidth}{%
    \smallskip
    \textbf{Key Takeaways (RQ3):}
    The analyses of the results we obtained for RQ3 show that using the One-shot with CoT prompting strategy is the best prompting strategy to automatically review assurance cases.
    }}

\subsection{(RQ4): LLM Generating the Best Automated Assurance Case Reviews}

 We also relied on Table \ref{tab:llm_comparison}
 to answer RQ4. This Table shows that DeepSeek-R1 is the most performant of all the LLMs at hand. It is closely followed by GPT 4.1. Both DeepSeek-R1 and GPT 4.1 therefore seem to have better reasoning capabilities when compared to GPT-4o and Gemini 2.0 Flash. This is a bit unexpected because, as seen in Table \ref{tab:LLM overview}, both GPT 4.1 and Gemini 2.0 Flash have the same cutoff date, which gave them the possibility to be trained on quite similar amounts of data. 
 On the other hand, as reported in Table \ref{tab:LLM overview}, GPT-4o has the earliest cutoff date. This could explain  its poorer performance compared to both GPT-4o and DeepSeek-R1.

\begin{table}[ht]
    \centering
    \renewcommand{\arraystretch}{1.1}
    \begin{tabular}{|l|l|c|c|c|c|c|}
        \hline
        \textbf{LLM} & \textbf{Strat.} & \textbf{AV A. C.} & \textbf{AV W-F} & \textbf{AV E. S.}  & \textbf{AV Arg. Cri.}  & \textbf{Gen AV}\\\hline
        GPT-4o & ZS & 3.82 & 3.2 & 3.6 & 3.73 & 3.59\\\cline{2-7}
               & ZS with CoT & 2.48 & 2.08 & 2.15 & 2.58 & 2.33\\\cline{2-7}
               & OS with CoT & 2.53 & 2.48 & 2.48 & 2.45 & 2.49\\\hline
        GPT 4.1 & ZS & 3.35 & 2.88 & 3.07 & 2.97 & 3.07\\\cline{2-7}
                & ZS with CoT & 2.4 & 1.63 & 2.38 & 2.35 & 2.19\\\cline{2-7}
                & OS with CoT & 2.15 & 1.87 & 2.52 & 2.22 & 2.19\\\hline
        DeepSeek-R1 & ZS & 3.15 & 2.7 & 2.83 & 2.98 & 2.92\\\cline{2-7}
                    & ZS with CoT & 2.27 & 1.55 & 2.25 & 2 & 2.02\\\cline{2-7}
                    & OS with CoT & 2.02 & 1.57 & 2.12 & 2.08 & 1.95 \\\hline
        Gemini 2.0 Flash & ZS & 3.22 & 3.27 & 3.25 & 3.53 & 3.32\\\cline{2-7}
                         & ZS with CoT & 2.67 & 2.38 & 2.68 & 2.6 & 2.58\\\cline{2-7}
                         & OS with CoT & 2.5 & 2.58 & 2.67 & 2.37 & 2.53\\\hline
    \end{tabular}
    \caption{Comparison of LLMs Results  Across the Four LLMs, Three Prompting Strategies, Five Runs, and Three Metrics (ZS = Zero-shot; OS = one-shot; Gen AV=General Average) 
        }
    \label{tab:llm_comparison}
\end{table}

\smallskip
\noindent\fbox{%
    \parbox{\linewidth}{%
    \smallskip
    \textbf{Key Takeaways (RQ4):}
    The analyses of the results show that DeepSeek-R1 is the best LLM when reviewing assurance cases, with 
    GPT 4.1 trailing slightly behind. Contrariwise, GPT-4o and Gemini 2.0 Flash are the least performant of the four LLMs under analysis.}}

\subsection{(RQ5): A Taxonomy  of LLM Review Capabilities and Issues}
\label{subsec:taxonomy}

We followed an approach similar to that of \cite{viger2024ai} to iteratively and incrementally create a taxonomy of LLM Review Capabilities and Issues. More specifically, the second assessor performed a qualitative analysis to semantically classify the LLM-generated reviews obtained in the three Experiments. This allowed the second assessor to iteratively create an initial classification of the review capabilities and the review issues characterizing the LLMs at hand. Another assessor (i.e. a senior co-author) reviewed that initial classification and met multiple times with the second assessor to refine the classification.
 Table \ref{tab:taxonomy} reports the resulting taxonomy. This taxonomy compiles a set of semantic topics and therefore provides a qualitative analysis that is complementary to the quantitative analysis we performed in Section \ref{sec: results}. The taxonomy allows us to detect the review capabilities and issues that the quantitative analysis may not have fully captured.
 
  Overall, all four LLMs exhibit quite similar review capabilities and issues. Still, some issues are more prevalent on some models, such as generating more generic responses when relying on Zero-shot prompts. This issue is more common with GPT-4o and GPT4.1 when compared to the DeepSeek-R1 and Gemini 2.0 Flash. More specifically, as Table \ref{tab:taxonomy} shows, each LLM can generate a detailed response for the CoT prompts. This, however, is only true for both DeepSeek-R1 and Gemini 2.0 Flash when relying on Zero-shot prompts, although Gemini 2.0 Flash also tends to give generic responses compared to DeepSeek-R1. Furthermore, each LLM is also able to identify any inconsistencies, spot some undeveloped goals, and identify ambiguous goals that are present in the assurance case. Another capability that we noticed is that some LLMs are able to detect if the GSN concepts used in an assurance case diverge from the ones supported by the GSN standard. For instance, LLMs can detect if a Context is not labeled as \textit{C}, or if a Justification is not labeled as \textit{J}. Specifically, GPT-4o, GPT 4.1, and DeepSeek-R1 can identify these issues when relying on the zero-shot prompts. Unexpectedly, the LLMs sometimes lose the capability to point this out as an issue when relying on the CoT prompts. Still, Gemini 2.0 Flash failed to demonstrate this review capability regardless of the prompting strategy at hand.

   \begin{table}[ht]
\centering
\begin{tabular}{|l|l|l|}
\hline
\textbf{Category} & \textbf{Topic} & \textbf{Corresponding LLMs} \\
\hline
\multirow{8}{*}{Capabilities} & Generation of a Detailed Response & All the LLMs \\
& Inconsistency identification  & All the LLMs \\
& Generation of concrete suggestions & GPT 4.1, DeepSeek-R1, Gemini 2.0 \\
& Identification of some undeveloped GSN elements & All the LLMs \\
& Identification of ambiguous goals & All the LLMs \\
& Identification of GSN concepts diverging from GSN & GPT-4o, GPT 4.1, DeepSeek-R1\\
\hline
\multirow{8}{*}{Issues} &Failure to use the specified  notation (CoT only)  & GPT-4o, GPT 4.1, DeepSeek-R1  \\
& Hallucination proneness  & All the LLMs \\
& Suggestion of responses that are too generic (ZS) & GPT-4o, GPT 4.1, Gemini 2.0  \\
& Failure to identify all identical GSN elements (WF) &All the LLMs \\
& Failure to pinpoint specific issues (ZS) & GPT-4o, GPT 4.1 \\
& Failure to identify some  undeveloped GSN elements &All the LLMs \\
\hline
\end{tabular}
\label{tab:taxonomy}
\caption{Taxonomy of Review Capabilities and Issues (WF= Well-formedness criterion; ZS= Zero-shot)}
\label{tab:taxonomy}
\end{table}
 
 The LLMs used in our experiments also exhibit some review issues. For instance, as Table \ref{tab:taxonomy} shows, most LLMs sometimes described the detected issues without using the mathematical notations we drove them to use through the prompts. This means that, sometimes, LLMs simply ignore the instructions specified in their prompts, which as \cite{shankar2024validates} pointed out, is a well-known issue inherent to LLMs. Interestingly, another issue that we noticed when reviewing the responses generated by the LLMs is that LLMs sometimes struggle to spot some of the undeveloped elements within the assurance cases they reviewed. Furthermore, each LLM, regardless of the prompting strategy, is prone to hallucinations. Our analysis also suggests that LLMs are sometimes unable to consistently identify all the duplicated GSN elements that are present in an analyzed assurance case. This therefore resulted in incomplete reviews that did not sufficiently meet the well-formedness criterion. Lastly, the other issue that we observed, although exclusively linked to the Zero-shot prompts, is that GPT-4o, GPT4.1, and Gemini 2.0 Flash sometimes generate reviews that are too generic and therefore lacking key insights. 
 
 The aforementioned review issues may lead to risks already identified in the literature (e.g., \cite{diemert2025one}), such as \textit{automation bias} (i.e., overreliance on LLMs to support a thorough review process), and the increase of assurance case development costs due to the need to manually refine and filter LLM-generated reviews. So, although our work highlights some promising LLM-based review capabilities, it also suggests that future work should focus on using more effective technologies to automate the review of assurance cases while making this task less risky and semi-automatic.
\smallskip
\noindent\fbox{%
    \parbox{\linewidth}{%
    \smallskip
    \textbf{Key Takeaways (RQ5):}
    Our taxonomy of LLM review capabilities and issues indicates that, while most LLMs exhibit quite similar review capabilities, DeepSeek-R1 and GPT 4.1 demonstrate superior review capabilities. Still, most of the LLMs at hand also exhibited some review issues. This suggests human intervention remains necessary to refine the reviews they generate and yield higher review quality.
    }}

\section{Threats to Validity} \label{sec:threats}

In the remainder of this Section, we rely on the framework that \cite{wohlin2012experimentation} proposed to discuss the threats to validity associated with our work.

\subsection{Internal Validity} 

A key threat to internal validity lies in the manual nature of our metric-based assessment, as no ground truth exists to benchmark LLM outputs. More specifically, the manual assessment of results (i.e. LLM-generated reviews) may result in ratings that are quite subjective and biased, since each reviewer may have different opinions regarding the correctness of reviews. Nevertheless, such disagreement is expected and typical in qualitative evaluation. 
To mitigate this threat, we involved two different reviewers in the review process, which led to the establishment of clear and consistent review guidelines to rate LLM-generated reviews.


\subsection{External Validity}
Because LLMs are trained on different datasets, their review results may vary accordingly. This may lead to different review results depending on the LLM at hand. This difference could hinder the generalization of our review results to other LLMs that were trained using different training data. Therefore, our findings should be interpreted in light of the specific LLMs evaluated. The generalization issue could also extend to other application domains, since there might not be enough training data on those other domains. As a result, LLMs may understand fewer contexts in these specific areas, leading to further generalization issues.

Furthermore, we randomly picked the assurance case we used as an example in our one-shot with CoT experiments. Still, as \cite{odu2024automatic} pointed out, the choice of an example may have an impact on the quality of the results these experiments yield, especially if the selected assurance case  focuses on a single application domain. Future work will explore the use of a systematic approach to better select examples for our experiments. 

\subsection{Construct Validity}
The way we formalized each review criterion may have influenced the quality of our experimental results by not sufficiently capturing the essence of these criteria. Future work will try to repeat the experiments by relying on refined predicates that may better formalize the review criteria at hand. These predicates may use more advanced mathematical symbols and notations that may better align with LLM learning capabilities. 

\subsection{Conclusion Validity}

The cutoff date of the LLMs, which indicates the data they were trained on, could threaten the validity of our results. In our context, the cut-off date is June 2024 for both Gemini 2.0 Flash and GPT-4.1, October 2023 for GPT-4o, and January 2025\footnote{Source: \url{https://github.com/HaoooWang/llm-knowledge-cutoff-dates}} for DeepSeek-R1.  These cutoff dates may affect LLM performance when completing the review task, as some assurance cases could overlap with training data. 
To mitigate this threat, we have therefore included some recent assurance cases (e.g., the safety case of Baidu Apollo) in our dataset. Still, no mitigation approach is entirely foolproof. So, despite reviewing relatively recent assurance cases, some residual risks may persist. For instance, the LLMs at hand could have encountered similar examples during training, or subtle biases might still remain.

\section{Conclusion} \label{sec:conclusion}
In this paper, we proposed a novel approach that allows us to determine if LLMs are capable of automatically reviewing assurance cases. Our experimental results show that, despite yielding promising results, LLMs are still unable to replace human experts when it comes to reviewing assurance cases. However, they can provide valuable support to human reviewers by highlighting potential issues or speeding up the review process. 

In future work, we will explore the use of additional prompting strategies in our experiments. These prompting strategies may include the \textbf{Persona} prompting strategy \citep{chen2024persona, olea2024evaluating, schulhoff2024prompt, kong2024better}. Also known as \textit{role prompting}, the Persona prompting strategy requires the LLM to "\textit{act}" as a person that has expertise on the field on which a task focuses, so that it provides more specific and noteworthy answers on the task. Relying on the persona prompt may allow us to create prompts that will make the LLM embody the personality of an assurance case reviewer to better perform the assurance case review.

Furthermore, when conducting our experiments, we relied on four different state-of-the-art LLMs specifically: Gemini 2.0 Flash, GPT-4o, GPT-4.1, and DeepSeek-R1. Although their results are promising, we believe it is crucial to also experiment with other LLMs such as Claude, LLaMA,  and Quem. The rationale is that, since different LLMs use different training datasets and support various reasoning capabilities, experimenting with additional LLMs may allow us to get very different and possibly richer insights regarding the ability of LLMs to support the automation of the assurance case review.

When assessing LLM-generated reviews, we observed that the LLMs sometimes flagged the supporting documents or artifacts mentioned in the assurance cases as being vague, probably because they were not fed to the LLMs as inputs. Based on these observations, we believe LLM-based assurance case review may be improved if the LLMs at hand are provided with additional context and can interact with the external environment. In this context, in future work, we plan to rely on an LLM-based agentic approach that will leverage  Retrieval Augmented Generation (RAG) to automatically support the assurance case review task. This LLM-based agentic approach may yield better results since it will rely on LLM-based agents that are capable of autonomously and iteratively improving their results thanks to their self-evolving capability \citep{zeng2025pilot, zhang2024survey, he2025llm, balu2025towards}. Integration of humans in the loop may be key to the success of this endeavor.

\section*{Data Availability}
\label{sec:data_availability}

Our replication package is available on this link: \url{https://doi.org/10.6084/m9.figshare.30369931.v4}

\section*{Acknowledgements}
We would like to thank Connected Minds and CFREF (Canada First Research Excellence Fund) for funding this study.

\section*{Appendices}
\label{sec:appendices}

\subsection{CoT Prompt Template for Well-Formedness (Syntax) }
\begin{table}[]
    \centering
    \caption{CoT Prompt Template for Well-Formedness (Syntax) } 
    \renewcommand{\arraystretch}{0.70}
    \begin{tabular}{p{7in}}
    \toprule
    Well-Formedness (Syntax) Prompt Template\\
    \midrule
    The Well-Formedness (Syntax) criterion focuses on determining if the assurance case has any structural errors and if there are any identical GSN elements. The key points on reviewing the Well-Formedness (Syntax) criterion are as follows:\\
    \\
    1. Determining if there are any disconnects in any of the GSN elements linkings which would cause misunderstandings.\\
    2. Determining if there are any structural errors in the assurance which includes circular arguments and arguments that always expects a specific answer.\\
    3. Determining if there are any GSN elements without supporting evidence that causes an incomplete understanding of the assurance case.\\
    4. Determining if there are any GSN labels that are identical to one another regardless of differing descriptions as each GSN label should be unique. \\
    \\
    To review the Well-Formedness (Syntax) criterion of an assurance case, there will be 6 subtasks that need to be followed and implemented accordingly, those 6 subtasks are as follows:\\
    \\
    \textbf{Listing out the GSN Labels} - Before proceeding to check on the Well-Formedness (Syntax) check, the first step is to take note of all the GSN labels and to do that listing out the GSN labels individually is crucial. Once listed out, group each GSN label under their own generalization (i.e. all G are listed under Goals, S under Strategies and so on).\\
    \\
    \textbf{Marking the identical GSN labels} - Once the GSN labels are listed out, identify any GSN labels that are repeated or are identical to any other GSN labels by explicitly marking them. This step is crucial since each GSN label should be unique and there should be no repetition of GSN labels. Make sure to carefully look at each GSN label and then mark them if there are any identical ones.\\
    \\
    \textbf{Identifying identical GSN labels} - The third subtask is to then identify any identical GSN labels. Using the marked GSN labels done in the previous steps, the following sequence notation will be used to let the user know about the identical GSN labels: <(\${E(n): The current GSN label}, \${IDEN(n): Textual description of the GSN element})>. In a scenario of 3 identical GSN label were found to contain G1, the notation thus would show <(\${E(1)}, \${IDEN(1)}, \${E(2)}, \${IDEN(2)}, \${E(3)}, \${IDEN(3)})> with which each E(n) and IDEN(n) consists of the GSN elements that were identified to be the ones with the identical GSN labels.\\
    \\
    \textbf{Identifying structural issues in the assurance case} - The next subtask is to identify any structural issues currently present in the assurance case. To denote any issues in the assurance case, the notation of Structural(\${E: The current GSN label}, \${D: Textual description of the GSN element}) will be used.\\
    \\
    \textbf{Description of issue} - Once the structural issues and identical labels are identified, the next step is to describe the issue based on the previous 2 subtasks and give an explanation to why it is an issue. To describe the issue in the assurance case, the notation of Description(\${I(n): The issue number}, \${E: The current GSN label}, \${ID: Description of the issue}) will be used.\\
    \\
    \textbf{Suggestion of Improvement} - Once the issue is described, the next step is to suggest an improvement to help solve the issue. To suggest an improvement, the notation of Suggest(\${I(n): The issue number}, \${E: The current GSN label}, \${Sg: Possible solution}) will be used.\\
    \\
    If there are no issues in the assurance case, the 6 subtasks can be skipped and instead of going through those subtasks, the score will just be issued along with a description of why the assurance case is perfect.\\
    \bottomrule
    \end{tabular}
    \label{tab:well-formedness}
\end{table}

\subsection{CoT Prompt Template for Expressive Sufficiency Checks}
\begin{table}[]
    \centering
    \caption{CoT Prompt Template for Expressive Sufficiency Checks}
    \begin{tabular}{p{6in}}
    \toprule
    Expressive Sufficiency Prompt Template\\
    \midrule
    The Expressive Sufficiency criterion focuses on determining if the assurance case elements are fully fleshed out. The key points on reviewing the Expressive Sufficiency criterion are as follows:\\
    \\
    1. Determine if there are any GSN elements that are ambiguous and need further explanation. Ambiguity can range from biases, prior knowledge or underlying meanings.\\
    2. Determine if there are any GSN elements with incomplete explanation or description.\\
    \\
    To review the Expressive Sufficiency criterion of an assurance case, there will be 3 subtasks that need to be followed and implemented accordingly, those 3 subtasks are as follows:\\
    \\
    \textbf{Identifying issues in the assurance case} - The first subtask is to identify any issues currently present in the assurance case. To denote any issues in the assurance case, the notation of Issue(\${E: The current GSN label}, \${D: Textual description of the current GSN element}) will be used.\\
    \\
    \textbf{Description of issue} - Once an issue is identified, the next step is to describe the issue and give an explanation to why it is an issue. To describe the issue in the assurance case, the notation of Description(\${I(n): The issue number}, \${E: The current GSN label}, \${ID: Description of the issue}) will be used.\\
    \\
    \textbf{Suggestion of Improvement} - Once the issue is described, the next step is to suggest an improvement to help solve the issue. To suggest an improvement, the notation of Suggest(\${I(n): The issue number}, \${E: The current GSN label}, \${Sg: Possible solution}) will be used.\\
    \\
    If there are no issues in the assurance case, the 3 subtasks can be skipped and instead of going through those subtasks, the score will just be issued along with a description of why the assurance case is perfect.\\
    \bottomrule
    \end{tabular}
    \label{tab:expressive sufficiency}
\end{table}

\subsection{CoT Prompt Template for Argument Criticism and Defeat}
\begin{table}[]
    \centering
    \caption{CoT Prompt Template for Argument Criticism and Defeat }
    \begin{tabular}{p{7in}}
    \toprule
    Argument Criticism and Defeat Prompt Template\\
    \midrule
    The Argument Criticism and Defeat criterion focuses on finding out any possible criticisms or flaws in the assurance case while also being able to determine any defeaters that are present in the assurance case. The key points on reviewing the Argument Criticism and Defeat criterion are as follows:\\
    \\
    1. Coverage: Determines to what extent the arguments presented cover the conclusion.\\
    2. Dependency: Determines if each GSN element is truly independent meaning changes to some GSN elements would not affect the other GSN elements.\\
    3. Definition: Determines if the GSN elements descriptor are over or under constrained meaning are the GSN elements trying to convey too much information that isn’t needed or conveying with too little information.\\
    4. Directness: Determines to what extent the supporting GSN elements relate back to the main goal.\\
    5. Relevance: Determines how relevant a supporting GSN element is to the main goal.\\
    6. Robustness: Determines the fragility of the supporting GSN elements to possible changes in the assurance case or to any of their consequent GSN elements.\\
    \\
    To review the Argument Criticism and Defeat criterion of an assurance case, there will be 4 subtasks that need to be followed and implemented accordingly, those 4 subtasks are as follows:\\
    \\
    \textbf{Identifying issues in the assurance case} - The first subtask is to identify any issues currently present in the assurance case. To denote any issues in the assurance case, the notation of Issue(\${E: The current GSN label}, \${D: Textual description of the current GSN element}) will be used.\\
    \\
    \textbf{Description of issue} - Once an issue is identified, the next step is to describe the issue and give an explanation to why it is an issue. To describe the issue in the assurance case, the notation of Description(\${I(n): The issue number}, \${E: The current GSN label}, \${ID: Description of the issue}) will be used.\\
    \\
    \textbf{Suggestion of Improvement} - Once the issue is described, the next step is to suggest an improvement to help solve the issue. To suggest an improvement, the notation of Suggest(\${I(n): The issue number}, \${E: The current GSN label}, \${Sg: Possible solution}) will be used.\\
    \\
    \textbf{Identifying Defeaters in the assurance case} - Once the issues have been determined and given possible solutions to them, the next sub task would be to identify possible defeaters that could harm the integrity of the assurance case. To identify any potential defeaters for the assurance case, the notation of Defeaters(\${D(n) The defeater number}, \${Df: The defeater}, \${De: The GSN label that will be defeated}) will be used.\\
    \\
    If there are no issues in the assurance case, the 4 subtasks can be skipped and instead of going through those subtasks, the score will just be issued along with a description of why the assurance case is perfect.\\
    \bottomrule
    \end{tabular}
    \label{tab:argument criticism}
\end{table}

\subsection{Zero-Shot System Prompt Template}
Table \ref{tab:zero-shot} describes the zero-shot system prompt template.

\begin{table}[h!]
    \centering
    \caption{Zero-Shot System Prompt Template}
    \begin{tabular}{p{7in}}
    \toprule
    Zero-Shot System Prompt Template\\
    \midrule
    You are an AI assistant who will assist in reviewing an assurance case represented using the Goal Structuring Notation (GSN). Your role as the assistant is to review the assurance case by scoring it using a linear scale ranging from 1 to 5, with 1 meaning the assurance case is totally correct and there is therefore no room for error and 5 meaning the assurance case is totally incorrect (i.e. full of errors). When the assurance case score is not 1, give a reasoning or give examples on how the assurance case can be further improved. More context about the assurance case will begin with the delimiter “@Context\_AC” and ends with the delimiter “@End\_Context\_AC” while the assurance case to be reviewed will begin with the delimiter “@Assurance\_Case” and end with the delimiter “@End\_Assurance\_Case”.\\
    \\
    \textcolor{red}{@Context\_AC}\\
    \\
    More Context Information on the assurance case to be placed here\\
    \\
    \textcolor{red}{@End\_Context\_AC}\\
    \\
    \textcolor{red}{@Assurance\_Case}\\
    \\
    The Assurance Case to be reviewed should be specified here in the structured prose format complying with GSN\\
    \\
    \textcolor{red}{@End\_Assurance\_Case}\\
    \\
    We have also defined which review criterion will be used for the review of an assurance case. This criterion will solely be used as a basis on how the assurance case should be reviewed and scored. The review criterion to be used will begin with the delimiter  “@Review\_Criterion” and end with the delimiter “@End\_Review\_Criterion” while the description of the review criterion will start with the delimiter “@Review\_Criterion\_Description” and end with the delimiter “@End\_Review\_Criterion\_Description”\\
    \\
    \textcolor{red}{@Review\_Criterion}\\
    \\
    Name of the review criterion to be specified here\\
    \\
    \textcolor{red}{@End\_Review\_Criterion}\\
    \\
    \textcolor{red}{@Review\_Criterion\_Description}\\
    \\
    Description of the review criterion to be placed here\\
    \\
    \textcolor{red}{@End\_Review\_Criterion\_Description}\\
    \\
    \bottomrule
    \end{tabular}
    \label{tab:zero-shot}
\end{table}

\subsection{Zero-Shot with CoT System Prompt Template }

Table \ref{tab:zero-shot-with-cot} describes the zero-shot with CoT system prompt template.

\begin{table}[h!]
    \centering
    \caption{Zero-Shot with CoT System Prompt Template }
    \renewcommand{\arraystretch}{0.85}
    \begin{tabular}{p{7in}}
    \toprule
    Zero-Shot with CoT System Prompt Template\\
    \midrule
    You are an AI assistant who will assist in reviewing an assurance case represented using the Goal Structuring Notation (GSN). Your role as the assistant is to review the assurance case by scoring it using a linear scale ranging from 1 to 5, with 1 meaning the assurance case is totally correct and there is therefore no room for error and 5 meaning the assurance case is totally incorrect (i.e. full of errors). When the assurance case score is not 1, give a reasoning or give examples on how the assurance case can be further improved. More context about the assurance case will begin with the delimiter “@Context\_AC” and ends with the delimiter “@End\_Context\_AC” while the assurance case to be reviewed will begin with the delimiter “@Assurance\_Case” and end with the delimiter “@End\_Assurance\_Case”.\\
    \\
    \textcolor{red}{@Context\_AC}\\
    \\
    More Context Information on the assurance case to be placed here\\
    \\
    \textcolor{red}{@End\_Context\_AC}\\
    \\
    \textcolor{red}{@Assurance\_Case}\\
    \\
    The Assurance Case to be reviewed should be specified here in the structured prose format complying with GSN\\
    \\
    \textcolor{red}{@End\_Assurance\_Case}\\
    \\
    We have also defined which review criterion will be used for the review of an assurance case. This criterion will solely be used as a basis on how the assurance case should be reviewed and scored. The review criterion to be used will begin with the delimiter  “@Review\_Criterion” and end with the delimiter “@End\_Review\_Criterion” while the description of the review criterion will start with the delimiter “@Review\_Criterion\_Description” and end with the delimiter “@End\_Review\_Criterion\_Description” with the chain of thought process on how the review should be done incrementally would begin with the delimiter “@Chain\_Of\_Thought” and end with the delimiter “@End\_Chain\_Of\_Thought”.\\
    \\
    \textcolor{red}{@Review\_Criterion}\\
    \\
    Name of the review criterion to be specified here\\
    \\
    \textcolor{red}{@End\_Review\_Criterion}\\
    \\
    \textcolor{red}{@Review\_Criterion\_Description}\\
    \\
    Description of the review criterion to be placed here\\
    \\
    \textcolor{red}{@End\_Review\_Criterion\_Description}\\
    \\
    \textcolor{red}{@Chain\_Of\_Thought}\\
    \\
    Chain of Thought text to be added here\\
    \\
    \textcolor{red}{@End\_Chain\_Of\_Thought}\\
    \bottomrule
    \end{tabular}
    \label{tab:zero-shot-with-cot}
\end{table}

\subsection{Contextual Information Supplied in the Prompts}

\begin{table}[h]
    \centering
    \caption{Context information (from the work of \cite{sivakumar2024prompting})}
    \begin{tabular}{p{6in}}
    \toprule
    Context used for LLM input \\
    \midrule
    An assurance case, such as a safety case or security case, can be represented using Goal Structuring Notation (GSN), a visual representation that presents the elements of an assurance case in a tree structure. The main elements of a GSN assurance case include Goals, Strategies, Solutions (evidence), Contexts, Assumptions, and Justifications. Additionally, an assurance case in GSN may include an undeveloped element decorator, represented as a hollow diamond placed at the bottom center of a goal or strategy element. This indicates that a particular line of argument for the goal or strategy has not been fully developed and needs to be further developed. I will explain each element of an assurance case in GSN so you can generate it efficiently. \\
    1. Goal – A goal is represented by a rectangle and denoted as G. It represents the claims made in the argument. Goals should contain only claims. For the top-level claim, it should contain the most fundamental objective of the entire assurance case. \\
    2. Strategy – A strategy is represented by a parallelogram and denoted as S. It describes the reasoning that connects the parent goals and their supporting goals. A Strategy should only summarize the argument approach. The text in a strategy element is usually preceded by phrases such as ‘‘Argument by appeal to. .. ’’, ‘‘Argument by .. .’’, ‘‘Argument across .. .’’ etc. \\
    3. Solution – A solution is represented by a circle and denoted as Sn. A solution element makes no claims but are simply references to evidence that provides support to a claim. \\
    4. Context (Rounded rectangles) – In GSN, context is represented by a rounded rectangle and denoted as C. The context element provides additional background information for an argument and the scope for a goal or strategy within an assurance case. \\
    5. Assumption – An assumption element is represented by an oval with the letter ‘A’ at the top- or bottom-right. It presents an intentionally unsubstantiated statement accepted as true within an assurance case. It is denoted by A.\\
    6. Justification (Ovals) – A justification element is represented by an oval with the letter ‘J’ at the top- or bottom-right. It presents a statement of reasoning or rationale within an assurance case. It is denoted by J.\\
    \bottomrule
    \end{tabular}
    \label{tab:context}
\end{table}

Table \ref{tab:context} reports the context information we used in our prompts.

\subsection{Review Criterion Descriptions Supplied in the Prompts}
Table \ref{tab:review_description} reports the description of each of the review criterion we focused on. We supplied these descriptions in our prompts.

\begin{table}[h]
    \centering
    \caption{Review Criterion description}
    \begin{tabular}{p{6in}}
    \toprule
    Review Criterion Descriptions\\
    \midrule
    \textbf{Argument Comprehension}: In this process, the reviewer must be able to identify key points of an argument which may include but not limited to claims, strategies, context of the argument, and also evidence of why the argument is valid.\\
    \textbf{Well-Formedness (Syntax)}: Finding out and identifying the structural errors in each argument. The reviewer should also be able to identify the claims on each argument without the need of using any supporting documents.\\
    \textbf{Expressive Sufficiency}: Ensures that arguments for the assurance cases are fully fleshed out so that the reviewer would be able to understand the argument clearly and will have no question regarding it.\\
    \textbf{Argument Criticism and Defeat}: This criterion focuses on finding possible criticisms that are present in the arguments and list out the defeaters that are present in the arguments.\\
    \bottomrule
    \end{tabular}
    \label{tab:review_description}
\end{table}

\subsection{GPCA Safety Case in the GSN format}
\begin{figure}
\centering
\includegraphics[width=0.75\linewidth]{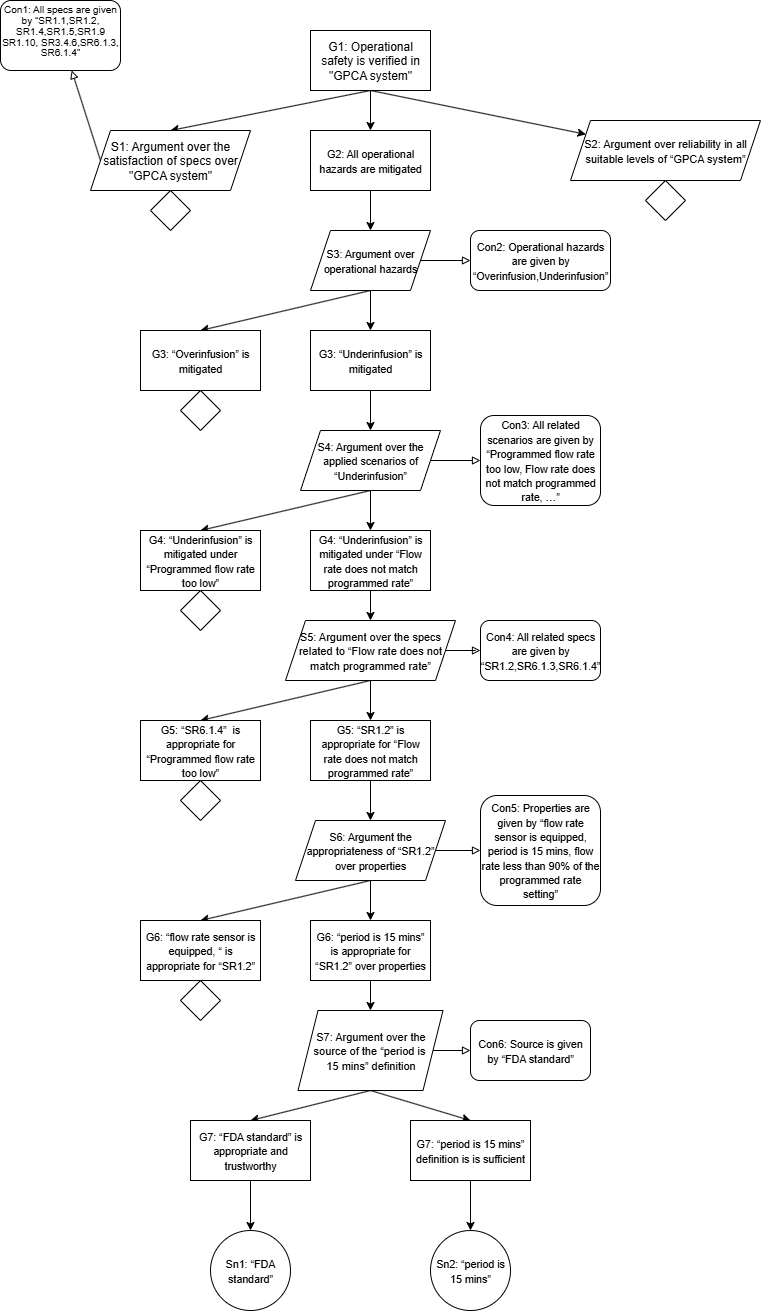}
\caption{A partial safety case of GPCA - adapted from \cite{lin2017support} }
\label{fig:GPCA Assurance Graphical}
\end{figure}

Figure \ref{fig:GPCA Assurance Graphical} illustrates the safety case of GPCA.

\subsection{Baidu Apollo Assurance Case Graphical Notation}
Safety Case of Baidu Apollo - adapted from \cite{odu2025llm}
\footnote{Github of Baidu Apollo Assurance Case: {\tiny \url{https://github.com/GerhardYu/LLMs-as-Judges-Toward-The-Automatic-Review-of-GSN-compliant-Assurance-Cases-Figures.git}}}

\bibliographystyle{elsarticle-harv} 
\bibliography{jss_paper_references}

\end{document}